\documentclass[twocolumn]{aastex631}

\usepackage{xcolor} 
\usepackage{amsmath,siunitx}
\usepackage{relsize} 
\usepackage{float} 
\usepackage{placeins} 
\usepackage{pifont} 
\usepackage{hyperref}
\usepackage{booktabs} 

\usepackage{graphicx}
\newcommand\T{\rule{0pt}{2.6ex}}
\newcommand\B{\rule[-1.2ex]{0pt}{0pt}}

\DeclareOldFontCommand{\rm}{\normalfont\rmfamily}{\mathrm}
\DeclareOldFontCommand{\sf}{\normalfont\sffamily}{\mathsf}
\DeclareOldFontCommand{\tt}{\normalfont\ttfamily}{\mathtt}

\newcommand{\psr}{B1055 }
\newcommand{\xmm}{\textit{XMM-Newton} }
\newcommand{\pn}{EPIC-pn }

\newcommand{\sig}{$\sigma$ }

\newcommand{\nudot}{\dot{\nu}}

\definecolor{fgn}{rgb}{0.13, 0.55, 0.13}
\definecolor{pplum}{rgb}{0.44, 0.11, 0.11}
\definecolor{dca}{rgb}{0.64, 0.0, 0.0}
\definecolor{darkmagenta}{rgb}{0.55, 0.0, 0.55}
\definecolor{darkcerulean}{rgb}{0.03, 0.27, 0.49}
\definecolor{blgr}{rgb}{0.0, 0.87, 0.87}

\begin{document}

\title[B1055-52]{Multiwavelength Pulsations and Surface Temperature Distribution in the Middle-Aged Pulsar B1055--52.}

\correspondingauthor{Armin Vahdat}
\email{mv.armin@gmail.com}

\author[0000-0002-4026-5885]{Armin Vahdat}
\affiliation{Institut f\"ur Astronomie und Astrophysik, Universit\"at Tübingen, Sand 1, D-72076 T\"ubingen, Germany}

\author[0000-0003-2317-9747]{B. Posselt}
\affiliation{Department of Astrophysics, University of Oxford, Denys Wilkinson Building, Keble Road, Oxford OX1 3RH, UK}


\author[0000-0002-7481-5259]{G.G. Pavlov}
\affiliation{Department of Astronomy \& Astrophysics, Pennsylvania State University, 525 Davey Lab, 16802 University Park, PA, USA}

\author[0000-0003-2122-4540]{P.~Weltevrede}
\affiliation{Jodrell Bank Centre for Astrophysics, Department of Physics and Astronomy, University of Manchester, Manchester M13 9PL, UK}


\author[0000-0003-4187-9560]{A. Santangelo}
\affiliation{Institut f\"ur Astronomie und Astrophysik, Universit\"at Tübingen, Sand 1, D-72076 T\"ubingen, Germany}

\author[0000-0002-7122-4963]{S.~Johnston}
\affiliation{CSIRO Astronomy and Space Science, Australia Telescope National Facility, PO~Box~76, Epping NSW~1710, Australia}




\begin{abstract}
We present a detailed study of the X-ray emission from PSR B1055--52 using \xmm observations from 2019 and 2000. The phase-integrated  X-ray emission from this pulsar is poorly described by existing neutron star atmosphere models. Instead, we confirm that, similar to other middle-aged pulsars, the best-fitting spectral model consists of two blackbody components, with substantially different temperatures and emitting areas, and a nonthermal component characterized by a power law. Our phase-resolved X-ray spectral analysis using this three-component model reveals variations in the thermal emission parameters with the pulsar's rotational phase. These variations suggest a nonuniform temperature distribution across the neutron star's surface, including the cold thermal component and  probable hot spot(s). Such a temperature distribution can be caused by external and internal heating processes, likely a combination thereof. We observe very high pulse fractions, 60\%--80\% in the $0.7-1.5$\,keV range, dominated by the hot blackbody component. This could be related to temperature non-uniformity and potential beaming effects in an atmosphere. We find indication of a second hot spot that appears at lower energies ($0.15-0.3$\,keV) than the first hot spot ($0.5-1.5$\,keV) in the X-ray light curves, and is offset by about half a rotation period. This finding aligns with the nearly orthogonal rotator geometry suggested by radio observations of this interpulse pulsar. If the hot spots are associated with polar caps, a possible explanation for their temperature asymmetry could be an offset magnetic dipole and/or an additional toroidal magnetic field component in the neutron star crust.
\end{abstract}



\section{Introduction} \label{sec:intro}

PSR B1055--52 (\psr hereafter, also known as PSR J1057--5226) is a middle-aged isolated neutron star (NS) sharing common observational properties with Geminga and PSR B0656+14. These three pulsars were dubbed the ``Three Musketeers'' by \cite{Becker1997} due to having similar spin-down energy loss rates, inferred surface dipole magnetic fields, distances, $\gamma$-ray emission and X-ray spectral properties. The X-ray emission of these pulsars includes thermal and nonthermal components. 
It is commonly accepted that most of the thermal radiation emanating from the bulk surface of these pulsars results from the heat transfer from the NS interior, hot spots are caused either by anisotropic heat transfer or returning currents from the magnetosphere or both, while the nonthermal emission is due to synchrotron radiation in the magnetosphere (e.g., \citealt{Pavlov2002, Harding1998ApJ}).

The spin frequency and the frequency derivative of \psr at the reference epoch MJD 57600 are $\nu$ = 5.07318862008(2) Hz and $\dot{\nu} = -150.272(1) \times 10^{-15}\;{\rm s}^{-2}$ where 1$\sigma$ uncertainty of the last significant digit is provided in parentheses \citep{Jankowski2019MNRAS}. This corresponds to the characteristic age $\tau$ = 535 kyrs, rotational energy loss rate $\dot{E} = 3.0 \times 10^{34} \:\mathrm{erg} \mathrm{s}^{-1}$, and 
surface dipole magnetic field strength $B = 1.1 \times 10^{12} \:\mathrm{G}$.

The source was discovered as a radio pulsar by
\citet{Vaughan1972}. The X-ray emission was discovered 
with the Einstein observatory \citep{Cheng1983}. The X-ray pulsations were detected with ROSAT by \citet{Oegelman1993}, who suggested that the X-ray spectrum consists of two components, including soft thermal emission from the NS surface. The results were confirmed with ASCA observations (\citealt{Becker1997}, and references therein).

Chandra observations of B1055 revealed that its X-ray spectrum, like those of the other two Musketeers, is best described by a three-component model, comprising cold and hot blackbodies as well as a power-law component  \citep{Pavlov2002}. \xmm observations of the Three Musketeers allowed for a phase-resolved, low-resolution X-ray spectroscopy which highlighted some differences between the three Musketeers in surface temperature distribution, and orientations of the rotation and magnetic axes \citep{DeLuca2005ApJ}.
There are also differences regarding the X-ray detection of pulsar wind nebulae (PWNe) around these close ($\lesssim 500$\,pc) NSs. \psr and PSR B0656+14 have surprisingly faint PWNe while the three-tail Geminga's PWN is relatively bright (e.g., \citealt{caraveo2003geminga, de2006complex, Posselt2015, Birzan2016, Posselt2017}).

In radio, B1055 exhibits a unique emission profile, featuring a main pulse (MP) and an interpulse (IP) separated by approximately $160^\circ$ in pulse longitude \citep{Weltevrede2009MNRAS}. These components likely originate from distinct magnetic poles. The MP-IP separation remains consistent across different radio frequencies, suggesting a highly inclined magnetic axis relative to the pulsar's rotation axis (\citealt{Weltevrede2009MNRAS}, and references therein).

In the $\gamma$-ray regime, B1055 exhibits an unusual profile, as emphasized in the Third Fermi-LAT pulsar catalog (3PC; \citealt{ThirdPC}). It has three  overlapping principal peaks with indications of two small peaks preceding and following the outer steep wing of the principal $\gamma$-ray emission components.  Only two of the 150 young pulsars in the 3PC show such a complex profile, and the origin of these profiles is not understood. 

The distance to \psr is still a matter of debate. The ATNF Pulsar Catalogue (ver.\ 1.68) reports 0.093 kpc, based on the DM-based value (DM = 29.69$\pm$0.01, \citealt{Petroff2013}) and the electron density model of \cite{Yao2017ApJ}, while a distance of 0.714 kpc is derived using the NE2001 model \citep{Cordes2002}.
\cite{Mignani2010} suggested a much smaller distance of 0.2--0.5 kpc by investigating the contributions of the individual thermal components to the multiwavelength spectrum. This distance can also explain the unusually high $\gamma$-ray efficiency of the pulsar that has been previously obtained assuming a distance of 750 pc \citep{Abdo2010ApJS}.
Throughout this paper, we will scale the distance to 0.35 kpc to calculate the luminosity and other related quantities.

In this paper we report the results of two new \xmm observations of \psr carried out in 2019 (PI B.\ Posselt). \citet{Posselt2023arXiv} analyzed the phase-integrated data for a long-term flux variability, but did not find significant differences between these new and the 2000 \xmm data. 
However, the new data allow for a better characterization of the spectral and timing properties of the source and tighter constraints on the geometry of the pulsar.\\

The paper is structured as follows. In Section \ref{sec:obsdat}, we describe the data and our data reduction procedure. Section \ref{sec:phint} is dedicated to phase-integrated spectroscopy, for which we combine the previous 2000 \xmm data with our new observations to better constrain temperatures and test various atmosphere models. 
Timing analysis results are reported in Section \ref{sec:timing}, where we discuss the pulse profile and its properties such as pulsed fraction and phase-energy map. In Section \ref{sec:phres} we focus on phase-resolved spectroscopy of B1055. Lastly, in Section \ref{sec:disc} we discuss our results.

\section{Observations and Data Reduction}
\label{sec:obsdat}
\subsection{X-ray data}
The  \xmm observatory observed \psr 
on 2019 June 21 for 85 ks and 2019 July 9 for 81 ks (Table \ref{tab:b1055xmm}). During both observations, the European Photon Imaging Camera (EPIC) \citep{Str2001AA} was operated in small-window (SW) mode for \pn (time resolution 5.7 ms), while EPIC-MOS1 and EPIC-MOS2 cameras \citep{Turner2001AA} were operated in full-frame (FF) mode (time resolution 2.6 s). To better constrain the phase-integrated spectral fit parameters (Section \ref{sec:phint}), we also reanalyzed the previous \xmm observations of \psr taken on on 2000 December 14 and 15 for 24 ks and 57 ks, respectively \citep{DeLuca2005ApJ}. 

Unfortunately, due to the the collapse of data along the CCD columns, the EPIC-pn data of 2000 obtained in Timing mode, are faced with significantly higher background contamination resulting in a lower signal-to-noise ratio (S/N) above a few keV compared to the 2019 observations.

The data processing of the four EPIC observations was done with the \xmm Science Analysis System (SAS) ver.\ 20.0.0 \citep{Gabriel2004} applying standard tasks. The dead-time corrected net exposure times, corresponding camera modes and filters, and count rates are given in Table \ref{tab:b1055xmm}.

\begin{table*}
\centering \caption{\xmm EPIC observations of B1055 used in this study}
\label{tab:b1055xmm}

\footnotesize
\begin{tabular}{lcccccccccccc}
\toprule
\midrule
Obs. ID & Start time  & \multicolumn{3}{c}{Net Exposure time (ks)} & \multicolumn{3}{c}{Modes and filters} & \multicolumn{3}{c}{Total count rate (Net source count fraction, \%)} \\[3pt]
        & (MJD) & pn & MOS1 & MOS2 & pn & MOS1 & MOS2 & pn & MOS1 & MOS2 \\[3pt]
\midrule
0113050101 & 51892 & 19.3 & 21.0  & 21.0 & Ti-ME & FF-ME & FF-ME & 444.7 $\pm$ 6.2 (83) & 141.7 $\pm$ 2.6 (98) & 147.7 $\pm$ 2.7 (99)\\
0113050201 & 51893 & 51.1 & 53.4  & 53.4 & Ti-ME & FF-ME & FF-ME & 435.8 $\pm$ 3.8 (82) & 147.5 $\pm$ 1.7 (99) & 148.9 $\pm$ 1.7 (98)\\
0842820101 & 58654 & 51.7 & 75.5  & 75.0 & SW-TN & FF-ME & FF-ME & 764.2 $\pm$ 3.9 (97) & 124.7 $\pm$ 1.3 (96) & 114.4 $\pm$ 1.3 (96)\\
0842820201 & 58673 & 54.4 & 77.7 & 77.7 & SW-TN & FF-ME & FF-ME & 738.2 $\pm$ 3.7 (97) & 120.4 $\pm$ 1.3 (96) & 121.3 $\pm$ 1.3 (96)\\
\bottomrule\\[-5pt]
\end{tabular}
\renewcommand{\thefootnote}{\fnsymbol{footnote}}\tablecomments{Ti-ME: timing mode with medium filter, SW-TN: small window with thin filter, FF-ME: full frame with medium filter. Total count rate (cts ks$^{-1}$) in $0.3-10$\,keV for 2019 observations and $0.4-10$\,keV for 2000 observations.}
\raggedright
\end{table*}

\begin{figure}
    \includegraphics[width=0.95\columnwidth]{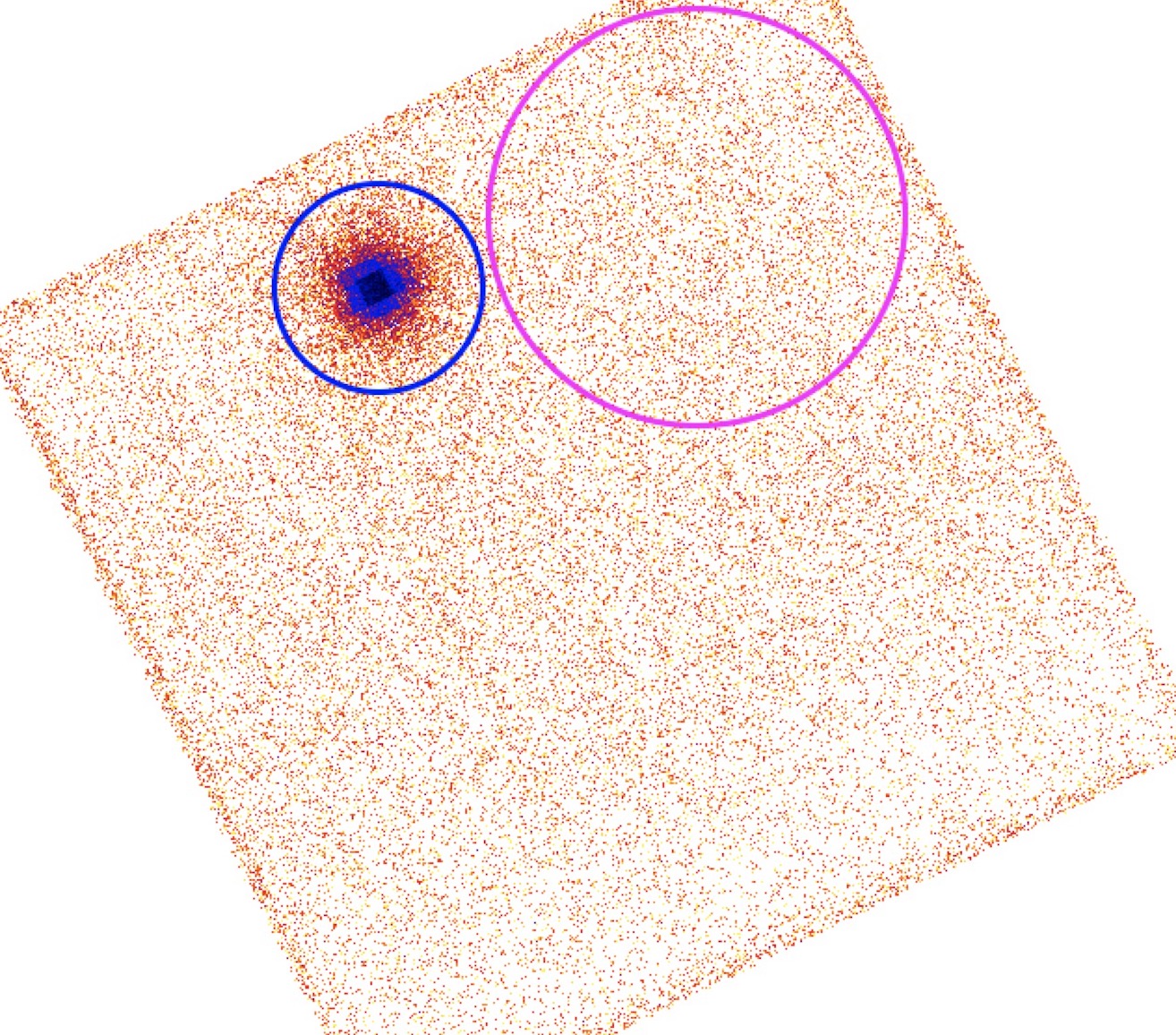}
    \includegraphics[width=0.95\columnwidth]{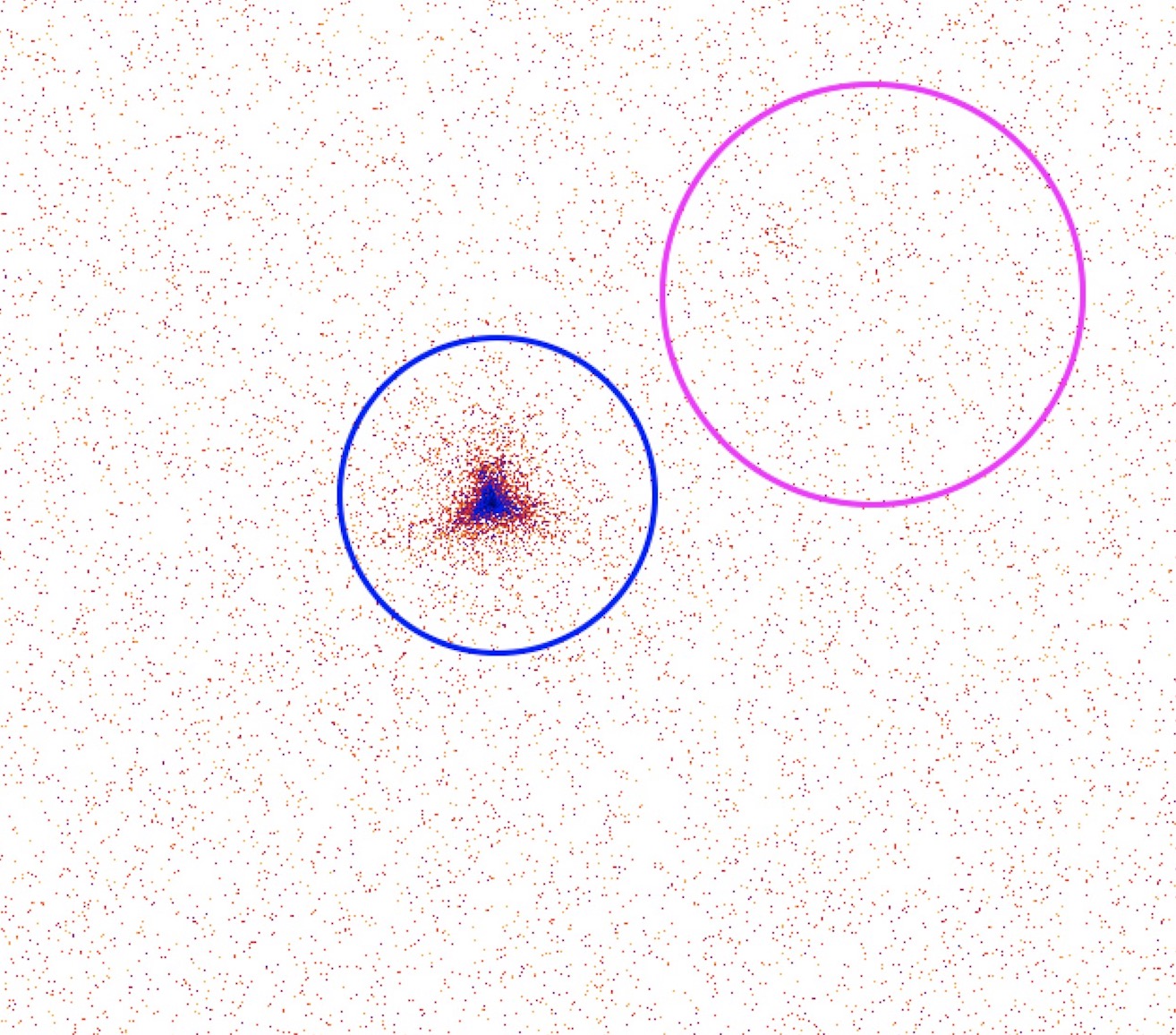}
    \caption{Count map of the field in the direction of the \psr for the first observation. The source extraction regions of $\ang{;;30}$ (EPIC-pn, top) and $\ang{;;45}$ (EPIC-MOS2, bottom) radii are displayed in blue whereas the background region of $\ang{;;60}$ radii is shown with magenta color. The images are smoothed with a Gaussian kernel with $\sigma=1\farcs25.$}.
    \label{fig:ds9}
\end{figure}

\subsection{Radio data}
We obtained radio pulsar data simultaneously with the \xmm observations. Two 34\,ks observations (program ID P1010) were carried out on MJD 58\,655 and MJD 58\,673 with the CSIRO Parkes $64$\,m radio telescope (also known as \emph{Murriyang}). We used the UWL receiver (0.7--4.0\,GHz) and the MEDUSA backend (e.g., \citealt{Hobbs2020}).
The observations were excised of radio frequency interference (RFI) and calibrated using standard \texttt{PSRCHIVE} tools \citep{Hotan2004,vanStraten2012}. From the averaged pulse profile, we produced a smoothed, high S/N template. By cross-correlating this template in the Fourier-domain \citep{Taylor1992}, we generated pulse times of arrival (ToAs) for 57 subintegrations for each Parkes observation.\\

We also obtained simultaneous and contemporaneous data on \psr from the MeerTime Thousand Pulsar Array (TPA) programme \citep{Johnston2020}. The observations were carried out with the 64-dish SARAO MeerKAT radio telescope. The TPA used the L-band receiver (centred at a frequency of 
1284\,MHz) and a total bandwidth of 775\,MHz. Details on the data reduction and RFI removal can be found in \citet{Parthasarathy2021, Lazarus2016}, details on the TPA procedures (such as template production) are reported by \citet{Posselt2023}.
We used three observing epochs: MJD\,58\,655.65 (simultaneous with \xmm, 5.4\,ks),  MJD\,58\,678.68 (202\,s), MJD\,58\,689.72 (202\,s) to generate ToAs with the appropriate MeerKAT-based template.

\subsection{\texorpdfstring{$\gamma$}--ray data}

We used the Fermi Large Area Telescope (LAT) $\gamma$-ray data set for this pulsar that was provided with the 3rd Fermi-LAT Pulsar Catalog (hereafter 3PC; \citealt{ThirdPC}). The data set consists of photons detected between MJD 54682 and 58791, including the epochs of our X-ray and radio observations, with energies between 50 MeV and 300 GeV.\\

\section{Phase-integrated Spectral Analysis}
\label{sec:phint} 

\subsection{Spectral extraction}
\label{subsec:specext} 

\xmm observations  of 2000 December 14--15 were carried out in Timing mode for pn and Full Frame (imaging) mode for MOS1 and MOS2.  Following the works of \cite{Posselt2015,Posselt2023arXiv}, we extracted pn spectra covering energies higher than 0.4 keV since soft background noise can impact the extracted spectra below this energy. We used 33 $\leqslant$ RAWX pixel $\leqslant$ 39 to extract the source spectra of both observations (101 and 201 hereafter)  whereas 5 $\leqslant$ RAWX  $\leqslant$ 7 and 4 $\leqslant$ RAWX  $\leqslant$ 6 were used for background regions of 101 and 201, respectively.

The 2000 EPIC-MOS1/2 data were extracted in the same way as described in \cite{Posselt2015} (e.g., circular apertures of $\ang{;;45}$ and $\ang{;;60}$ radii for source and background regions)
 
The EPIC data of 2019 June 20 and July 19 were collected in imaging modes (see Table 1). For \pn the source region is a circle with a radius of $\ang{;;30}$. We utilized the SAS tool {\ttfamily eregionanalyse} to fine-tune the position and the aperture that maximizes the source-to-background count ratio. For EPIC-MOS1/2 we used the circular aperture of the $\ang{;;45}$ radius, similar to the data of 2000, to extract the phase-integrated spectra. As shown in Figure \ref{fig:ds9}, we have selected the background regions with a circular aperture of $\ang{;;60}$ radius, sufficiently close to the pulsar to get an accurate estimate of the nearby background contribution but far enough to avoid contamination from the point-spread function of the source.

\subsection{Spectral fitting}
\label{subsec:specfit}

We started our analysis by comparing the two observations of 2019 (101N and 201N hereafter) with each other. We have checked and confirmed the consistency of the two data sets in terms of source and background count ratio and distribution and spectral fit parameters.

We restricted our energy range to $0.3-8$\,keV. Above 8\,keV, not only the background contribution exceeds the source flux but also the accuracy of our fit parameters did not improve further with increasing the upper energy bound.

For X-ray spectral fitting, we made use of PyXspec ver.\ 2.1.0 python interface with XSPEC ver.\ 12.12.0 \citep{Arnaud1996, Gordon2021}.

We used the Tübingen-Boulder model through its XSPEC implementation {\ttfamily tbabs} to fit the interstellar absorption by setting the abundance table to {\ttfamily wilm} \citep{2000Wilms} and photoelectric cross-section table to {\ttfamily vern} \citep{Verner1996}.

We first fitted the high energy part ($2.3-8$\,keV) of the spectra with an absorbed power-law (PL) model ({\ttfamily powerlaw} in XSPEC). 
Although the absorbing hydrogen column density, $N_{\text H}$, generally does not affect the spectrum at high energies, we fixed $N_{\text H}$ and varied $N_{\text H}$ at different values and checked the confidence contours of PL-index ($\Gamma$)-normalization for the two epochs and found an agreement within 1\sig in each case. 

In the next step, we fit the $0.3-8$\,keV broad-band spectra using a three-component model. This model is comprised of two blackbody (BB) models ({\ttfamily bbodyrad} in XSPEC) and a PL model with a best-fit value of $\Gamma=1.8$, determined from the PL-only fit. We also allowed $N_{\text H}$ to vary freely. For the two observations of 2019, we compared the confidence contours of temperature-normalization for the hot and cold BB components and found that their values agreed within a $3\sigma$ range. 

\begin{figure}
	\includegraphics[width=1.0\columnwidth]{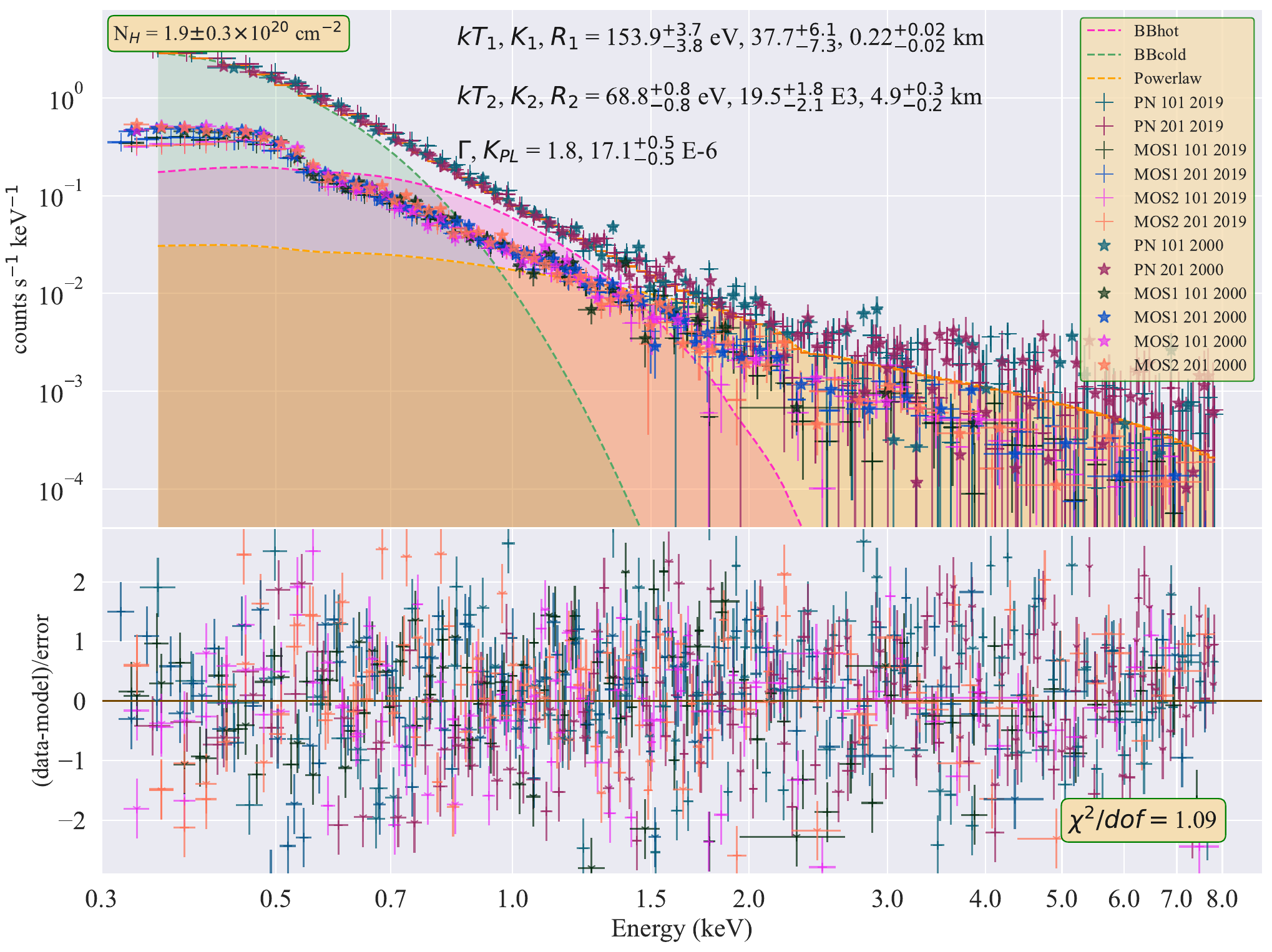}
    \caption{0.3-8 keV phase-integrated spectral fit to the 
    combined (2000 and 2019) data set of 12 spectra with 2BB+PL model (Table \ref{tbl-2_intspectra}). Hot BB, cold BB, and PL components are displayed with red, green, and yellow colors. The fit parameters with a margin of 1\sig error are also shown at top of the figure.}
    \label{fig:PIS_12}
\end{figure}

After establishing consistency, we combined the two data sets of 2019 and compared the combined data with the data of 2000 by repeating the same checks. While we obtained a similar consistency level between the 2000 and 2019 data as before, the count statistic and background noise level were different, which was most noticeable at high energies. 

Finally we combined the 12 spectra, and carried out a 2BB+PL fit. We obtained the best fit ($\chi^2_{\upsilon}$ = 1.09; where $\upsilon$=822 is the number of degrees of freedom) with BB temperatures $kT_1$ = 153.9$\pm$3.7 eV, $kT_2$ = 68.6$\pm$0.8 eV, and $N_{\text H}$ = (1.9 $\pm$ 0.3) $\times$ 10$^{20}$ cm$^{-2}$ (Figure \ref{fig:PIS_12} and Table \ref{tbl-2_intspectra}).

\begin{table}[th]
\centering
\caption{Parameters for the 2BB+PL fit of the phase-integrated spectrum in the $0.3-8$\,keV band for the combined 12 data sets. Errors are reported 
at the $1\sigma$ confidence level. $K_{\rm PL}$, $K_{\rm BB,c}$ and $K_{\rm BB,h}$ are \texttt{powerlaw}, cold and hot \texttt{bbodyrad} normalization parameters.
\label{tbl-2_intspectra}}
\begin{tabular}{lc}
\toprule
\midrule
Parameter (Units) & Best fit results \\
\midrule
$N_{\rm H}$ ($10^{20}$ cm$^{-2}$){\def\hfill{\hskip 50pt plus 1fill}\dotfill} & 1.9 $\pm$ 0.3  \\
\T\B
$kT_{\rm BB,c}$ (eV) \dotfill & $68.6 \pm 0.8$  \\
\T\B
$K_{\rm BB,c}$ ($10^{3}$ $R^2_{\rm km}$/$D^2_{\rm 10}$) \dotfill & $19.5 \pm 1.9$	  \\
\T\B
$R_{\rm BB,c}$ (km) \dotfill & $4.94 \pm 0.1$  \\
\T\B
$kT_{\rm BB,h}$ (eV) \dotfill & $153.9 \pm 3.7$ 	 \\
\T\B
$K_{\rm BB,h}$ ($R^2_{\rm km}$/$D^2_{\rm 10}$) \dotfill & $37.7 \pm 6.5$  \\
\T\B
$R_{\rm BB,h}$ (km) \dotfill & $0.22 \pm 0.02$  \\
\T\B
$\Gamma$ \dotfill & $1.8$  \\
\T\B 
$K_{\rm PL}$ $(10^{-6}$ photons/keV/cm$^{2}$/s at 1\,keV)
\dotfill & $17.1 \pm 0.5$ \\
\T\B\T\B
$\chi^2$/dof \dotfill & 899/822 \\
\T\B\T\B
$F^{\rm abs}_{0.3 - 8\,{\rm keV}}$ ($10^{-12}$ erg cm$^{-2}$ s$^{-1}$) \dotfill & $1.50 \pm 0.02 $ \\
\T\B\T\B
$L^{\rm 350 pc}_{0.3 - 8\,{\rm keV}}$ ($10^{31}$ erg s$^{-1})$ \dotfill & $2.80 \pm 0.12$ \\
\bottomrule\\[-4pt]
\end{tabular}
\end{table}

\begin{figure}
	\includegraphics[width=1.0\columnwidth]{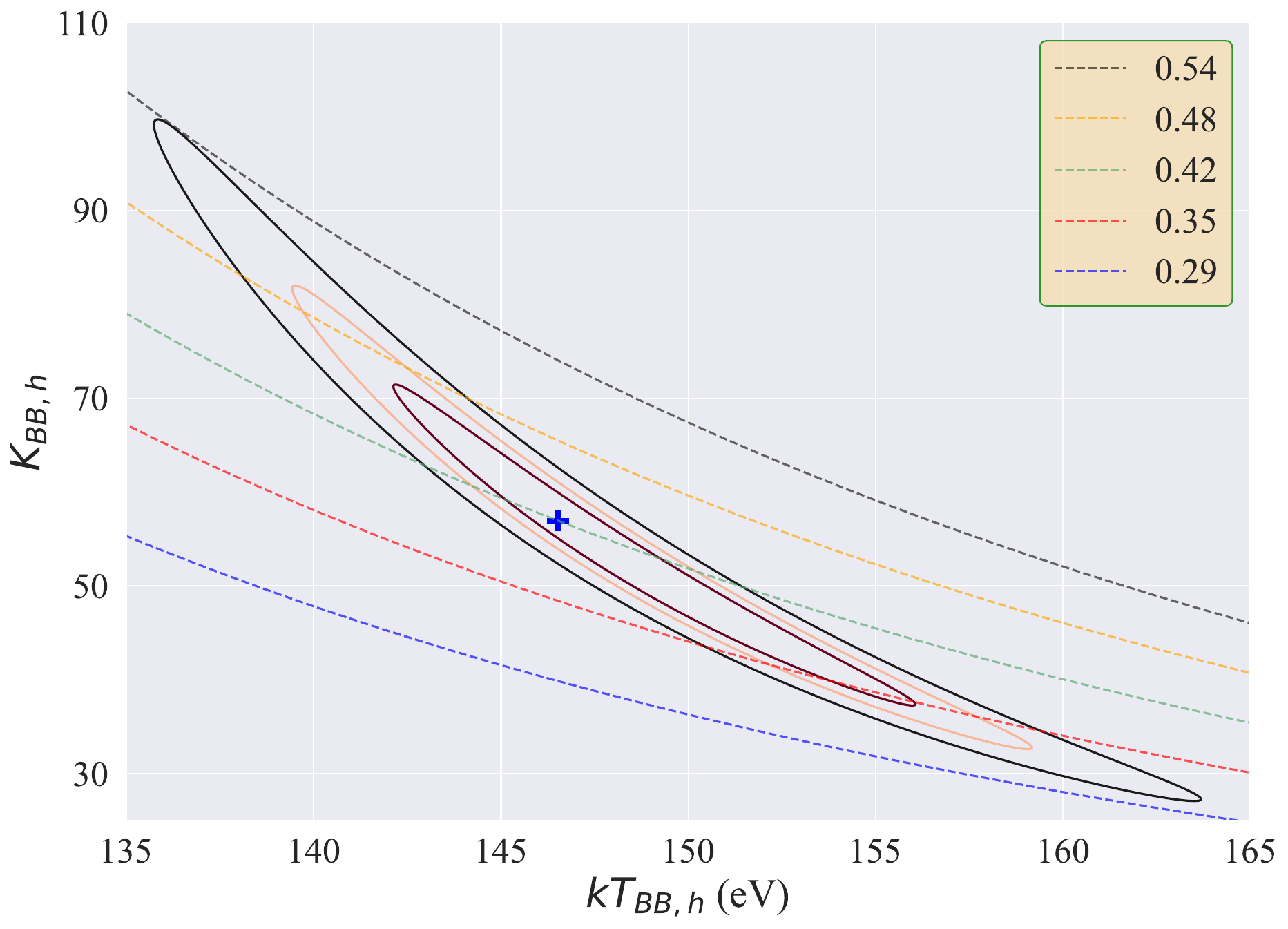}
	\includegraphics[width=1.0\columnwidth]{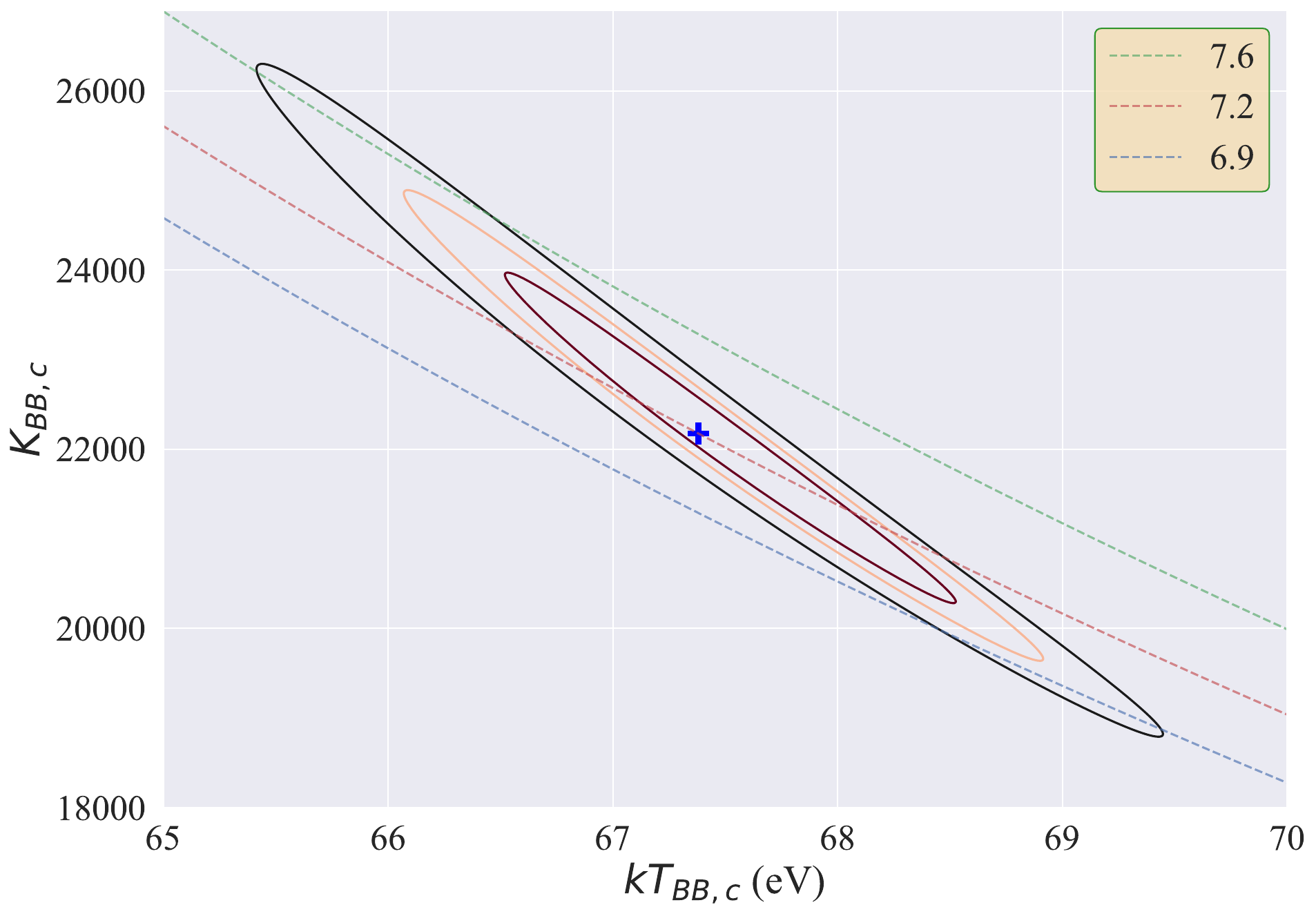}
    \caption{
    Hot BB (top) and cold BB (bottom) confidence contours for phase-integrated spectra using  101N and 201N pn data. Blue plus signs correspond to minimum values of $\chi^2$. Contours of constant bolometric luminosity  of an equivalent sphere,
    $L_{\rm BB} = 4\pi R_{\rm BB}^2\sigma T_{\rm BB}^4$, in units of $10^{31}$ erg s$^{-1}$ are overplotted in the $kT_{\rm BB}$-$K_{\rm BB}$ planes. The red,  orange and black contours correspond to confidence levels of 68.3\%, 90\%, and 99\%, respectively.}
    \label{fig:cont}
\end{figure}

We also tested several NS atmosphere models available in XSEPC and applicable to atmospheres with magnetic fields of about $10^{12}$ G: NSATMOS \citep{Heinke2006ApJ}, NSA (\citealt{Pavlov1995}; \citealt{Zavlin1996}), and NSMAXG
(\citealt{Mori2007}; \citealt{Ho2008ApJ}). Fitting with single-component and two-component atmosphere models resulted in large systematic residuals at higher energies ($\chi^2_{\upsilon} \gtrsim $  8) whereas two-component and three-component models (by adding a PL or BB) also produced unacceptable fits as a consequence of large systematic residuals at low energies, $\lesssim$ 1.5 keV. 

\section{Timing Analysis}
\label{sec:timing}
\subsection{Radio timing solutions}
We applied the pulsar-timing code, \texttt{tempo2} \citep{Hobbs2006} for our ToAs to determine the timing solutions for either the two Parkes or the three MeerKAT observations. For this, we used reference epoch MJD\,58\,664.376 (near the middle of the two X-ray observations of 2019,  see Section~\ref {subsec:timingsol}), a contemporaneous (MJD\,58\,602) \emph{Hubble} position and the proper motion from \citet{Posselt2023arXiv}.
From the \texttt{tempo2} fits, we obtain the frequency, $\nu$, and its time derivative, $\dot{\nu}$, see Table~\ref{tbl_radio}.
The difference in $\nu$ is $0.13$\,nHz between the Parkes and MeerKAT timing solutions. This difference is  $1.9\sigma$ of the MeerKAT timing solution, the difference $\dot{\nu}$ has a similarly low significance level ($1.3\sigma$). Hence, the two independent radio timing solutions agree with each other. For comparison with the X-ray data, we will use in the following the Parkes radio timing solution. 

\begin{table}
\centering
\caption{The radio timing properties of B1055.\label{tbl_radio}}
\begin{tabular}{cl}
\toprule
\midrule
\T\B
Parameter & Value \\
\tableline
\T\B
Right ascension (J2000)  \dotfill  &  10:57:59.0123\\
\T\B
Declination (J2000) \dotfill &   $-52$:26:56.509 \\
\T\B
Position epoch (MJD) \dotfill &	58602	\\
\T\B
Radio timing epoch (MJD) \dotfill	&	58664.376	\\
\T\B
Dispersion measure (DM) \dotfill    &	29.69 cm$^{-3}$ pc	\\
\tableline
\T\B
Parkes MJD time span  \dotfill	& $58655.096 - 58673.441$ \\
\T\B
Parkes Frequency ($\nu$) \dotfill	&	5.073174809812(17) Hz	\\
\T\B
Parkes Frequency derivative ($\dot\nu$) \dotfill &	$-1478(16)\times10^{-16}\;{\rm s}^{-2}$ \\

Parkes tempo2 RMS residual \dotfill & 17.6\,$\mu$s\\
\tableline
\T\B
MeerKAT MJD time span  \dotfill	& $58655.651 - 58689.720$ \\
\T\B
MeerKAT Frequency ($\nu$) \dotfill	&	5.073174809686(70) Hz	\\
\T\B
MeerKAT Frequency derivative ($\dot\nu$) \dotfill &	$-1500(1)\times10^{-16}\;{\rm s}^{-2}$ \\
MeerKAT tempo2 RMS residual \dotfill & 95.1\,$\mu$s\\
\bottomrule\\[-4pt]
\end{tabular}
\end{table}

\subsection{Relative \texorpdfstring{$\gamma$}--ray timing solution}

We folded this Fermi-LAT data for this pulsar using the 3PC ephemeris \citep{ThirdPC}, which was developed from 12 years of monitoring with Parkes. We replaced the ``TZRMJD'' and ``TZRFRQ'' parameters with those from our own radio timing ephemeris to phase-align the $\gamma$-ray pulse profile with our own radio profile, and found a radio/$\gamma$-ray alignment very similar to that found in 3PC.

\subsection{X-ray timing solutions} \label{subsec:timingsol}

Since the time resolution of the MOS detectors in FF mode is as large as 2.6 seconds, they are not suitable for the timing analysis of B1055. Therefore, we only use the 101N and 201N pn-observations which have a nominal frame time of 5.7 ms in SW mode. We corrected all event times in our data sets to the solar system barycenter using the standard SAS task {\ttfamily barycen}. 

The first event in 101N observation was detected at MJD 58654.849534 barycentric dynamical time; TDB (or $t_{1}$ = 677449468.950103 s in \xmm mission reference time; MRT). For the second observation, the first event 
was detected at MJD 58673.018533 TBD (or $t_{2}$ = 679019270.421713 s in MRT).

\vspace{1cm}
\begin{figure}
\includegraphics[width=1.0\columnwidth]{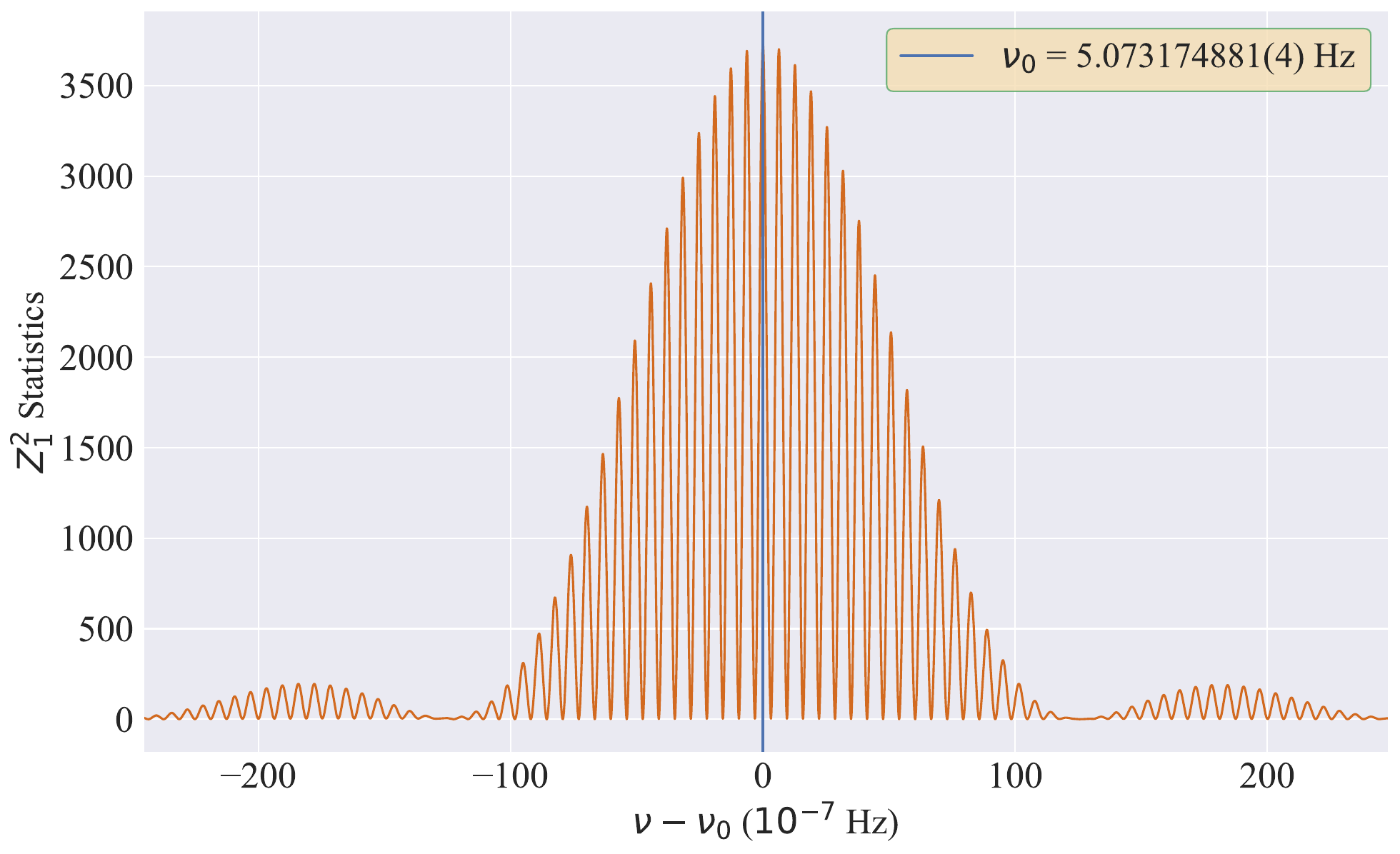}
    \includegraphics[width=1.0\columnwidth]{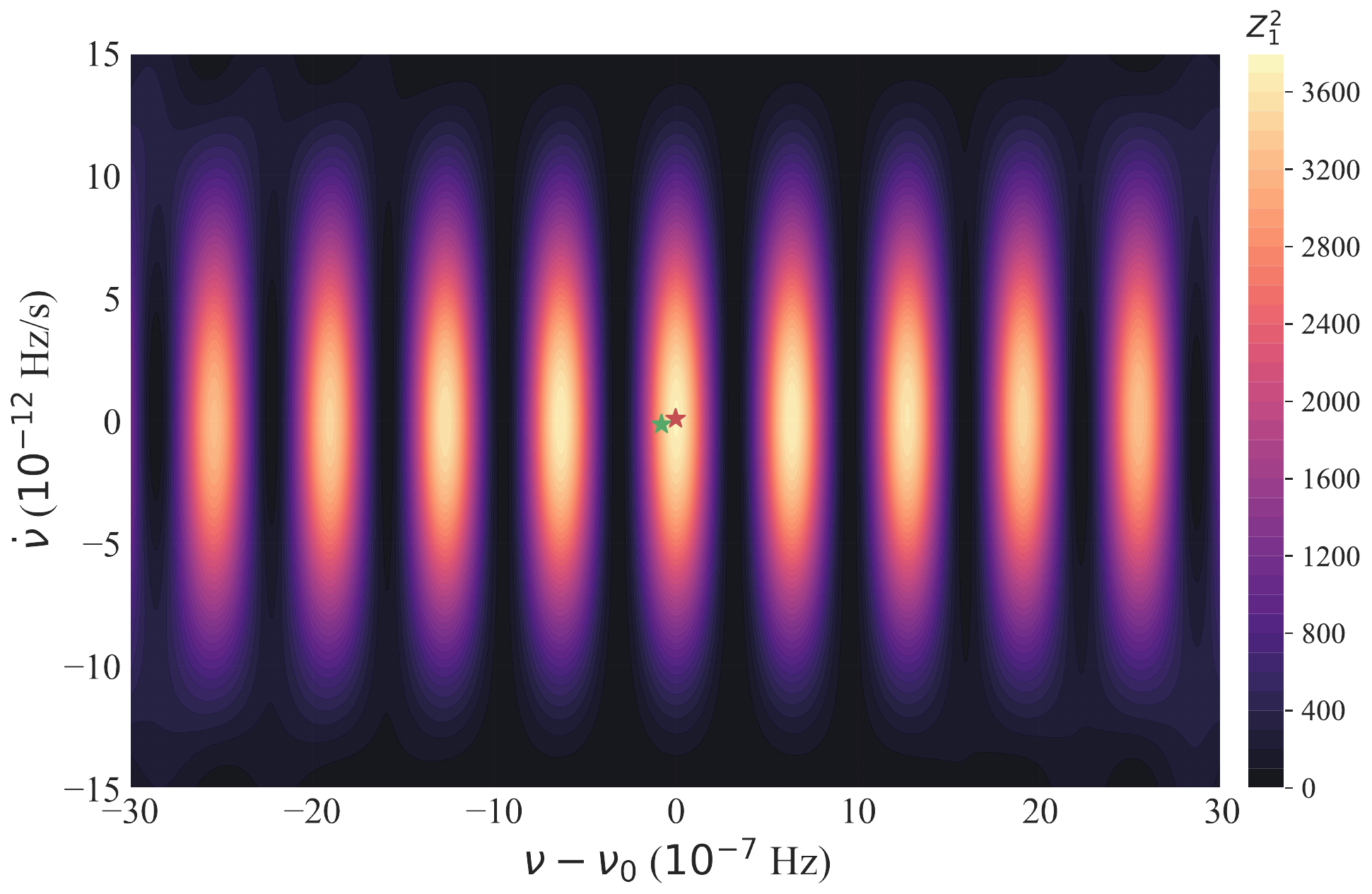}
    \includegraphics[width=1.0\columnwidth]{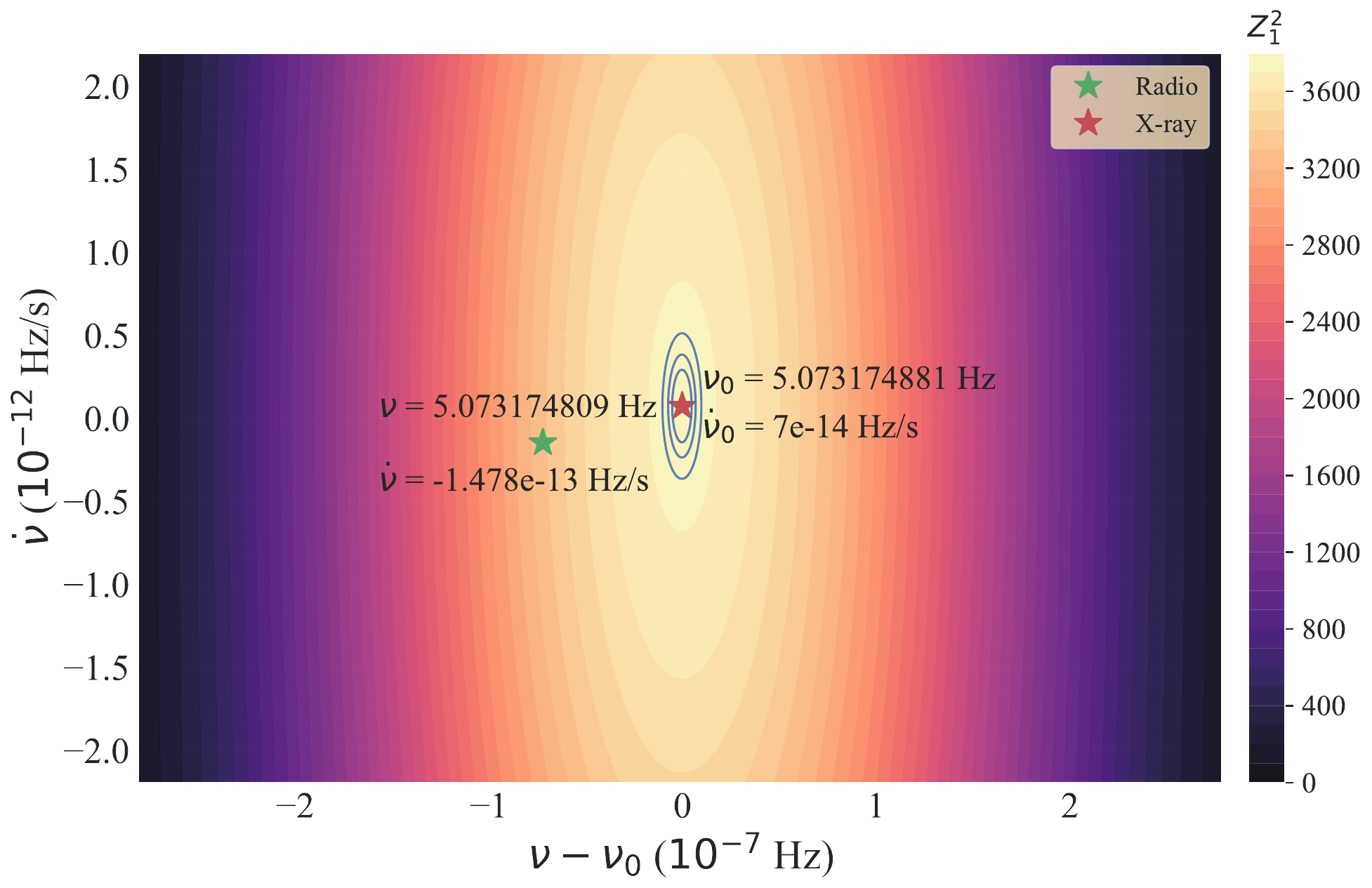}
    \caption{Top: $Z^2_{1}$ statistics with a fixed $\dot{\nu} = 7\times 10^{-14}$ Hz\,s$^{-1}$ obtained for $0.3-7$\,keV events extracted from EPIC-pn source region. Middle and bottom: Maps of $Z^2_{1} (\nu, \dot{\nu})$ in the vicinity of pulsar's expected (green star) frequency and its derivative from Parkes radio timing solution for the energy range $0.3-7$\,keV. The red star indicates the location of the highest $Z^2_{1} (\nu, \dot{\nu})$ value ($Z^2_{1, \rm max} = 3739$). The 1,2 and 3\sig statistical error contours for $Z^2_{1} (\nu, \dot{\nu})$ are displayed with blue ellipses.}
    \label{fig:2dmap}
\end{figure}

The times elapsed between the first and last detected events for 101N and 201N data sets were $T_{\mathrm{span1}} = 79988.797120$ s ($\approx 0.9258$ days) and $T_{\mathrm{span2}} = 79135.1205240$ s ($\approx0.9159$ days), respectively.

We used the $Z_n^2$ statistic \citep{Buccheri1983} to study X-ray pulsations in the vicinity of the frequency expected from the radio timing solution. 
We optimized GTI screenings, energy regions, and extraction apertures to ensure high signal-to-noise ($S/N$) extraction and large $Z^2_{n, \text {max}}$ values.
Prior to the $\nu$ and $\dot{\nu}$ search in our X-ray data, we subtracted ($t_{1} + T_{\mathrm{span2}} + t_{2})/2$ from the event times of 101N and 201N to minimize the correlation between $\nu$ and $\dot{\nu}$, so that the measured ephemeris corresponds to the middle of the two observations (i.e., $T_{\mathrm{ref}}$ = MJD 58664.376681).

We first performed the $Z^2_n$ test separately for the two data sets at fixed $\dot{\nu} = -1.478 \times 10^{-13}$ Hz\,s$^{-1}$ (the Parkes frequency derivative; see Table 3), using the \texttt{stingray} python package \citep{Huppenkothen2019ApJ}. For 101N (201N), we extracted 44113 (41984) events from GTI-filtered $0.3-7$\,keV data using the extraction region of a $\ang{;;30}$ radius, which provided the largest $Z^2_{n, \rm max}$ values. From the H-test \citep{deJager1989} we found that the H-statistic reaches a maximum for three harmonics ($n=3$) for each of the data sets. We obtained $Z^2_{1,\mathrm{max}}$ = 1925 (1802) and $Z^2_{3,\mathrm{max}}$ = 2123 (2029) for 101N (201N) data at the frequency $\nu = 5.07317468(13)$ Hz ($5.07317498(14)$ Hz).  The numbers in brackets indicate the $1\sigma$ uncertainties
calculated as: $Z_1^2(\nu_0\pm \sigma_\nu) = Z_{1, \rm max}^2 - 1$, where $Z_1^2(\nu_0) = Z_{1,\rm max}^2$.

The frequencies obtained from the two individual X-ray observations agree with each other within $2\sigma$, and within $5\sigma$ with the frequency measured by the Parkes telescope.

In the next step, we combined the two \pn data sets (86,097 events in the $0.3-7$\,keV band) and chose the middle of our two observations as the reference time (MJD 58664.376681). For the total time span $T_{\rm span(1+2)}\approx 19.085$ d, the combined observation should allow one to measure not only frequency, but also frequency derivative. Therefore, we calculated $Z_n^2$ on a $\nu$-$\dot{\nu}$ grid and found $Z^2_{1, \text {max}}$ = 3739 ($Z^2_{3, \text {max}}$ = 4155) at $\nu(Z_{1\rm max}^2) = 5.073174881(4)$ Hz and $\dot{\nu}(Z_{\text{1,max}}^2) = +7 \times 10^{-14}$ Hz\,s$^{-1}$ (see Figure \ref{fig:2dmap}). 

The $Z^2_{1, \text {max}}$ and $Z^2_{3, \text {max}}$ values in the summed data are very close to the sums of $Z^2_{1, \text {max}}$ and $Z^2_{3, \text {max}}$ in separate data sets (see Table \ref{tbl:zstat}), indicating that we reached a good phase connection between the two data sets and obtained a coherent timing solution. However, the difference of 72 nHz between the 
frequencies measured in the X-ray and Parkes radio data exceeds the formal statistical uncertainty of 4\,nHz by a factor of 18. The positive value of $\dot{\nu}(Z_{\text{1,max}}^2)$ is in conflict with the well constrained, negative radio timing value, likely indicating an issue with the accuracy of the X-ray timing solution. Since the two independent radio timing solutions agree with each other, we explored technical reasons for the deviating \xmm results and possibilities for correction. Firstly, we checked possible time jumps in our two X-ray time-series. Time jumps can occur due to the known temperature effect on the frame time of the EPIC-pn detector as well as `reset' errors of the counter clock \citep{Kirsch2004ESASP}. In 101N, no real-time jumps have been recognized by SAS and in 201N they were successfully corrected (we set the SAS environment parameter {\ttfamily SAS\_JUMP\_TOLERANCE} to a value of 44).\footnote{For more information on the SAS environment parameter, please refer to SAS documentation}

The second observation lacks some internal house-keeping data that would allow a detailed monitoring of the detector temperature. Without these additional data, SAS relies on approximations for correcting the temperature-based clock drifts, which, however, usually work very well. Because of the lack of these additional data, we have investigated in particular whether the second observation may cause faulty results. Unrecognized time jumps could in principle occur in ground station switches. \citet{Gotthelf2020} had an \xmm observation in the same instrument mode immediately after our second observation. Their results do not indicate any unusual timing offset or timing problems. Instead of the three ground stations in our observation, the data by \citet{Gotthelf2020} uses only two. We excluded in our observation those times of the additional ground station, suspecting that it may introduce some problem. However, this did not change the X-ray determined $\nu$ (which was different by only 2\,nHz from the full X-ray data solution). Thus, ground station switches in the second observation are unlikely to have introduced any timing error.\\

Comparing radio and X-ray determined pulsation periods of different pulsars in different \xmm instrument modes, \citet{Martin2012} reported a relative timing accuracy, $\triangle P / {P_{\rm radio}}$ of better than $10^{-8}$, where $\triangle P$ is the difference between the measured period. For our \psr observations, the relative error is $\triangle P /{P_{\rm radio}} = 1.4 \times 10^{-8}$. 
This value seems to be reasonably close to expectation. Together with our checks for any possible time jumps, we therefore conclude that we have to add a systematic uncertainty on the order of 60\,nHz to our statistical uncertainty of the determined X-ray frequency.

Similarly, the $\dot{\nu}_X$ has a larger uncertainty.  
Although the timing solution derived from X-rays is less accurate than the one obtained from radio, using this specific X-ray timing solution maximises the pulsed X-ray signal. Since the resulting errors in the photon phases are small (see Section~\ref{subsec:timingprop}), we employed the derived values of $\nu_X$ and $\dot{\nu}_X$ in the subsequent X-ray analyses to maximise statistical robustness and consistency in X-rays.

\begin{table}
\centering
\caption{Maximum values of $Z^2_{n}(\nu, \dot{\nu})$ for different energy ranges and numbers of harmonics ($n$). $N$ and \%bgd are the number of total counts and percentage of the average background counts in the source aperture of the combined and individual (101N \& 201N) data. $\chi^2_{\upsilon}$ characterizes the agreement between the binned histograms and the harmonic description of the phase-folded light curve in each band. \label{tbl:zstat}}
\begin{tabular}{l l c c c c c} 
\toprule
\midrule
\T\B
Energies & $N$ (\%bkg) &  $Z^{2}_1$  & $Z^{2}_2$ & 
$Z^{2}_3$ & $Z_{4}^2$ & $\chi^2_{\upsilon}$\\
(keV) & 
& & & & & \T\B\\
\tableline
\multicolumn{7}{c}{Combined $Z^{2}_n$ values} \T\B\\
$0.15-0.3$ & 81103 (6.6) & 300 & 422 & 440 & 450 & 1.2 \T \B \\
$0.3-0.45$ & 47549 (2.6) & 833 & 983 & 1002 & 1004 & 0.9 \T \B \\
$0.45-0.6$ & 20866 (1.7) & 1209 & 1260 & 1291 & 1294 & 1.0 \T \B \\
$0.6-0.8$ & 9181 (2.5) & 1817 & 1922 & 1932 & 1939 & 0.9 \T \B \\
$0.8-1.2$ & 4304 (7.5) & 1109 & 1197 & 1210 & 1214 & 1.0 \T \B \\
$1.2-2.0$ & 1740 (27.2) & 186 & 223 & 242 & 246 & 1.5 \T \B \\
$2.0-5.0$ & 1542 (61.5) & 20 & 26 & 39 & 33 & 1.9 \T \B \\
$0.3-7.0$ & 86097 (4.7) & 3739 & 4082 & 4155 & 4159 & 1.0 \T \B\\
\hline
\multicolumn{7}{c}{101N (top) \& 201N (bottom) individual $Z^{2}_n$ values} \T\B \T\B\\
$0.3-7.0$ & 44113 (4.6) & 1925 & 2092 & 2123 & 2125 & 1.0 \T \B\\
$0.3-7.0$ & 41984 (4.7) & 1802 & 1985 & 2029 & 2031 & 1.0 \T \B\\
\bottomrule\\[-6pt]
\end{tabular}
\end{table}

\subsection{Timing properties}
\label{subsec:timingprop}

\begin{figure}
	\includegraphics[width=1.0\columnwidth]{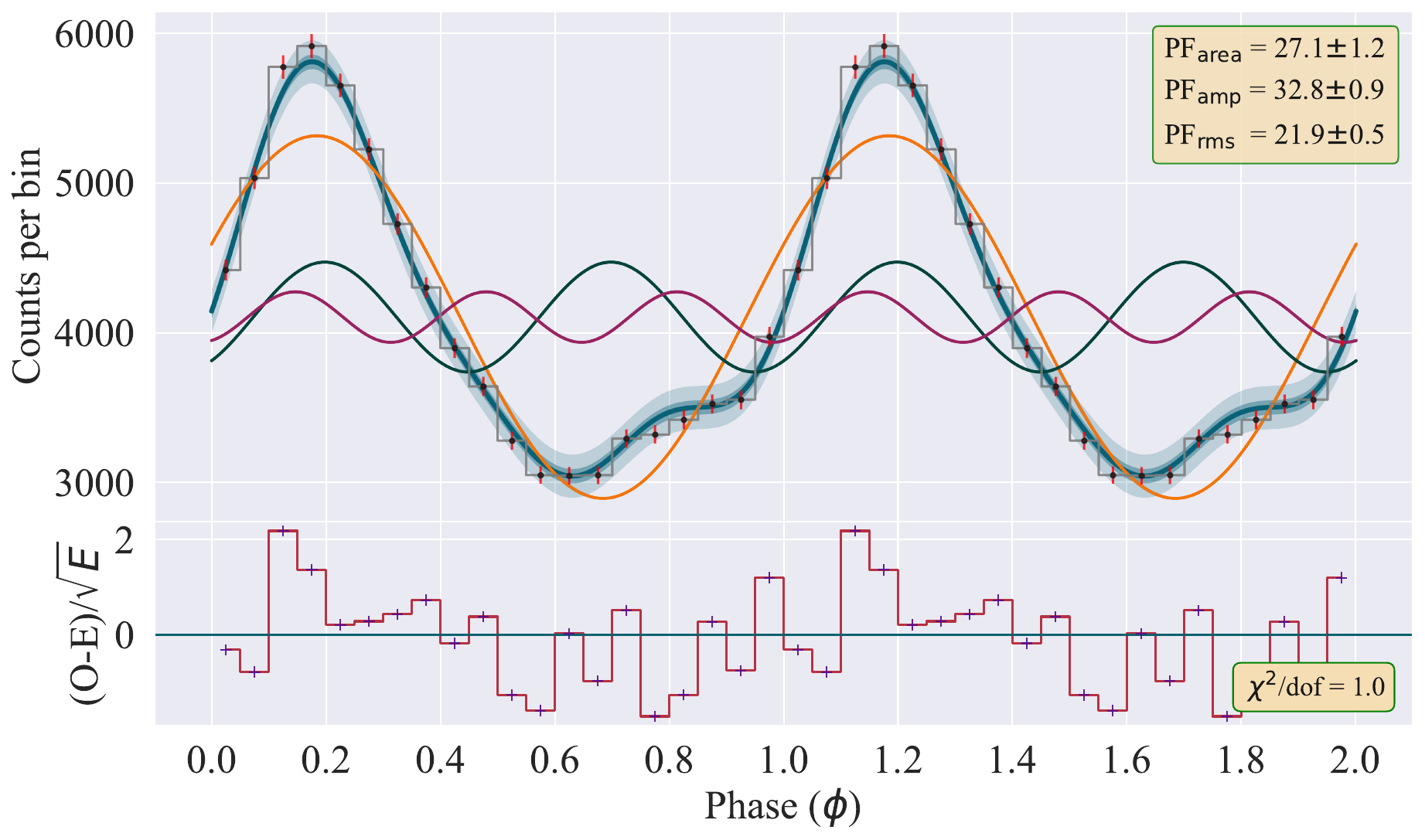}
    \caption{The $0.3-7$\,keV 
    phase-folded light curve (X-ray pulse profile) plotted as a histogram with 20 phase bins and as the sum of 3 harmonics, $(N/20){\cal F}(\phi)$, where ${\cal F}(\phi)$ is the normalized pulse profile and $N=86097$ is the total number of events in the chosen energy range. The $\pm3\sigma$ uncertainty of the X-ray pulse profile is shown with transparent blue color. The orange, green, and red sine waves correspond to the first, second, and third harmonics, respectively. Area, amplitude and rms pulse fractions in percents, with their $1\sigma$ uncertainties, are displayed at the top right corner.}
    \label{fig:pp}
\end{figure}

We studied the harmonic behavior of the X-ray pulse profile using the Fourier coefficients. For comparison, we also obtained the folded light curve in the form of a histogram in $0.3-7$\,keV with the timing solution determined above. The phase-folded light curve and contributions from each harmonic are displayed in Figure \ref{fig:pp}.
The appropriate number of harmonics can be different in different energy bands. We used a chi-square test from the Python library SciPy \citep{2020NatMe} to compare the sum of 3 harmonics with the 20-bin histogram of the folded light curve. 

The two descriptions of the pulse profile are in very good agreement. This result implies that binning is not necessary to examine the pulse profile in such a case of smooth pulsations, and additional uncertainties associated with the binning procedure can be avoided using the Fourier analysis.\\

\begin{figure}[t]
	\includegraphics[width=1.0\columnwidth]{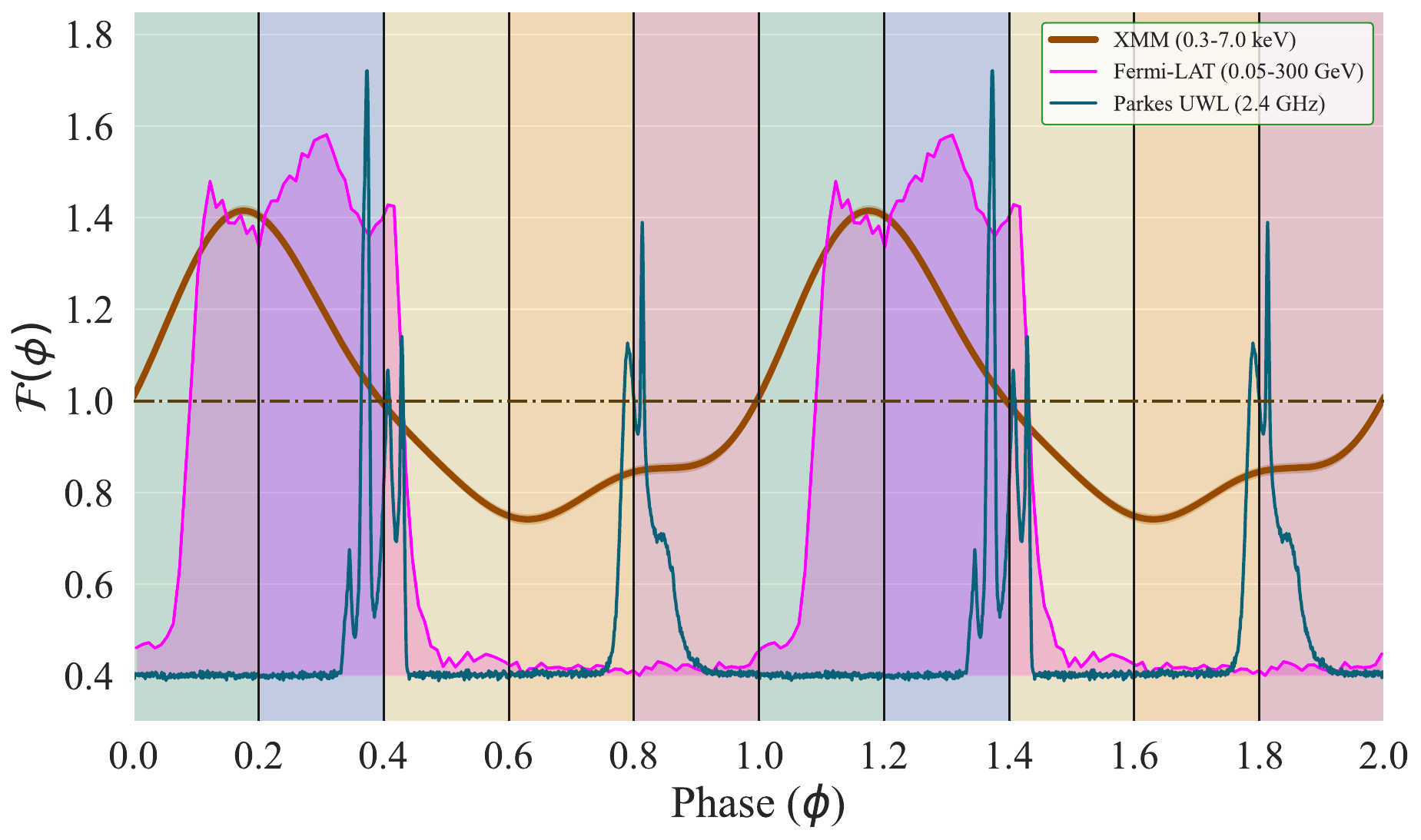}
    \caption{Normalized X-ray pulse profile, ${\cal F}(\phi)$, in $0.3-7$\,keV. Phase bins are displayed with different transparent colors. The Parkes radio light curve (at 2.4 GHz) and Fermi-LAT $\gamma$-ray light curve (0.05--300 GeV) are over-plotted  with blue and magenta colors, respectively.}
    \label{fig:pp_radio}
\end{figure}

\begin{figure}
	\includegraphics[width=1.0\columnwidth]{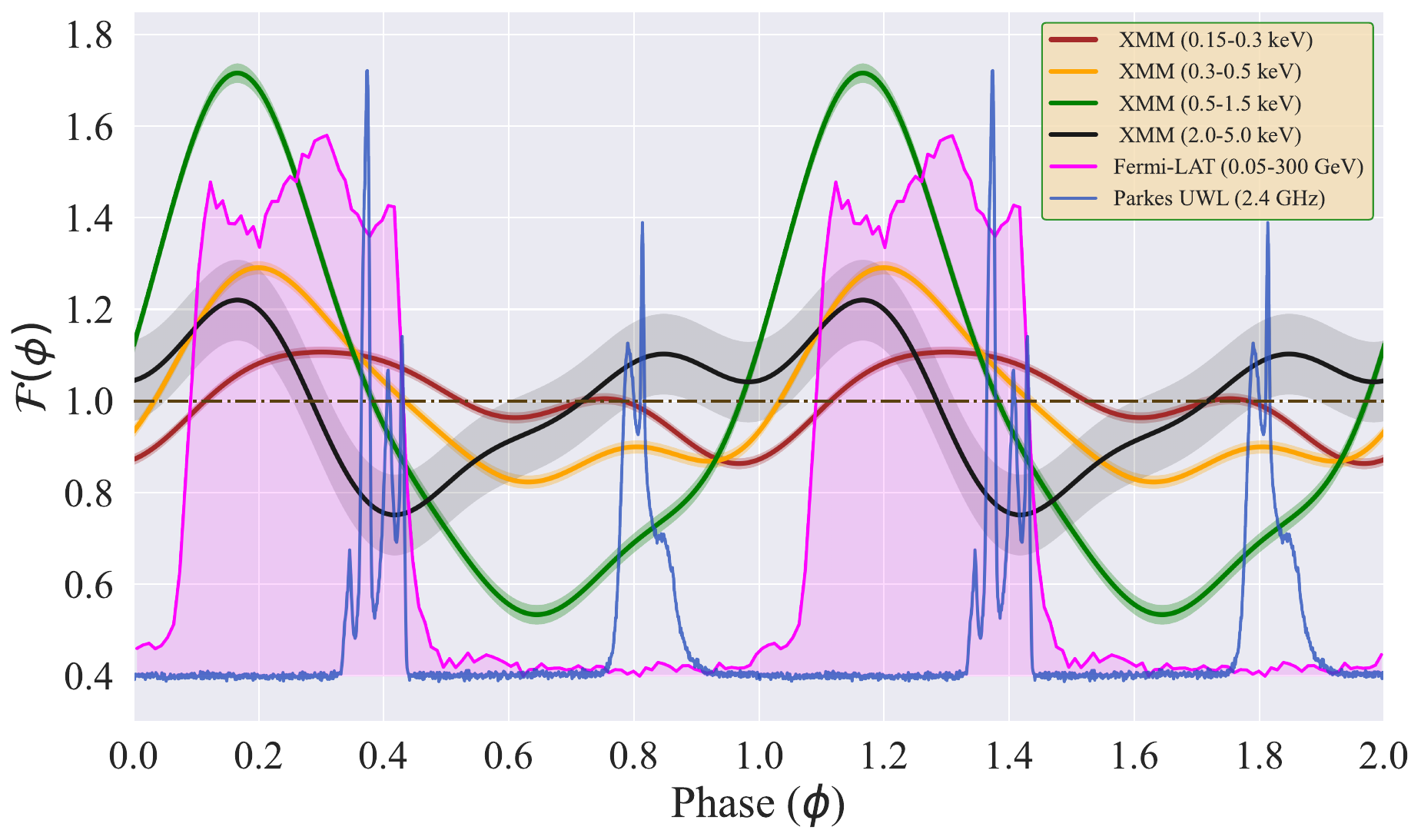}
    \caption{Normalized pulse profiles, ${\cal F}(\phi)$, are displayed for four energy bands. The X-ray pulse profiles' 1$\sigma$ uncertainties is depicted with transparent colors. The Parkes radio light curve (at 2.4 GHz) and Fermi-LAT $\gamma$-ray light curve (0.05--300 GeV) are over-plotted  with blue and magenta colors, respectively.}
    \label{fig:PR_5bin}
\end{figure}

The normalized X-ray pulse profile ${\cal F}(\phi)$ (see Equation (A1) in \citealt{Hare2021ApJ} for its definition), in the $0.3-7$\,keV range, is plotted over the Parkes radio profile at 2.4 GHz, and the Fermi-LAT $\gamma$-ray light curve from 50 MeV to 300 GeV is shown in Figure \ref{fig:pp_radio}. Figure \ref{fig:PR_5bin}  highlights the pulse profiles and their $3\sigma$ uncertainties in four energy bands together with Parkes radio and Fermi-LAT $\gamma$-ray light curves.\\

In order to estimate the intrinsic uncertainty in the alignment of the radio and X-ray profiles, we calculated the maximum phase shift, $\Delta\phi_{\rm max}$, between the X-ray timing solution ($\nu_X, \nudot_X$) and Parkes radio timing solution ($\nu_R, \nudot_R$) as follows.

For a given timing solution, the change of phase between times $t_1$ and $t_2$ (the first event in the first observation and the last event in the second observation) is:
\begin{equation}
    \Delta\phi = \left[\nu + \nudot\left(\frac{t_1+t_2}{2} - T_{\rm ref}\right)\right]\,\Delta t\,,
\end{equation}
where $\Delta t = t_2-t_1$ and $T_{\rm ref}$  is the reference time. The difference of phase changes between the radio solution and X-ray solution is:

\begin{eqnarray}
    \Delta\phi_X - \Delta\phi_R & = & (\nu_X-\nu_R) \Delta t \nonumber \\
    &+ & (\nudot_X-\nudot_R)\frac{t_1+t_2}{2} \Delta t \nonumber\\
    & -  &(\nudot_X T_{\rm ref,X} - \nudot_R T_{\rm ref,R}) \Delta t\,.
\end{eqnarray}
We estimated $\Delta\phi_{\rm max} = 0.12$, considering the midpoint of each observation as the reference times for the respective solutions, and $\Delta\phi_{\rm max} = 0.13$, considering the beginning of observation 1 and the end of observation 2 as the reference times.\\

\begin{figure}
	\includegraphics[width=1.0\columnwidth]{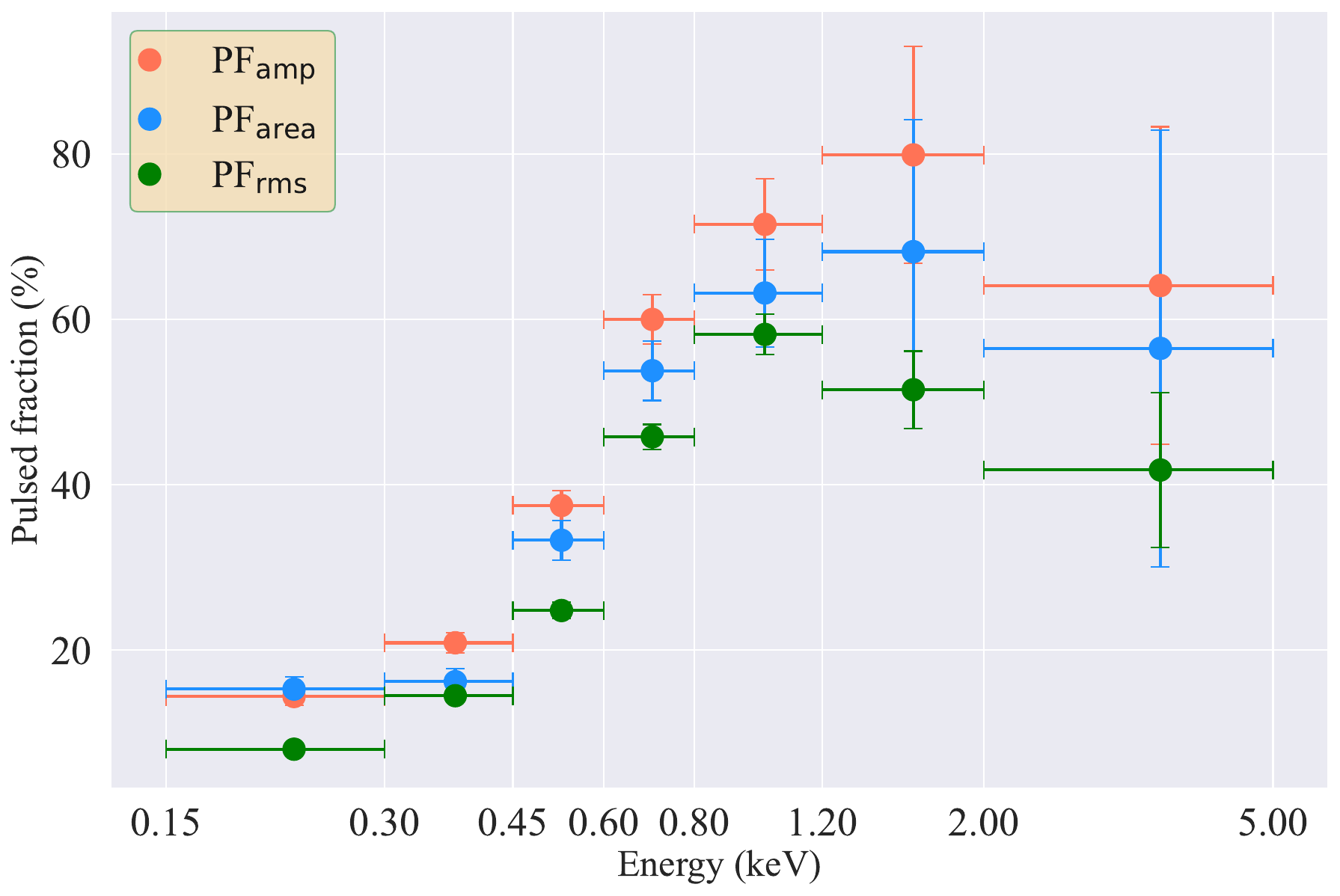}
    \caption{Energy dependence of three background-corrected pulse fractions plotted with 90\% confidence level errors for the energy ranges defined in Table \ref{tbl:zstat}. The orange, blue and green dots correspond to amplitude, area and root-mean-square pulse fractions.}
    \label{fig:pf}
\end{figure}

The pulsed fraction (PF) is one of the important properties of a folded light curve which primarily depends on the phases and amplitudes of Fourier harmonics and their dependence on energy. We estimated PFs for three different PF definitions, namely, the amplitude PF (${ p}_{\mathrm{amp}}$, sometimes referred to as ``peak-to-peak'' or ``max-to-min'' PF), the area PF (${p}_{\mathrm{area}}$), and the root-mean-square PF (${p}_{\mathrm{rms}}$; see Appendix C in \citealt{Hare2021ApJ} for exact definitions). The energy dependence of the PFs is shown in Figure \ref{fig:pf}.
Both the amplitude and area PFs increase with the photon energy, saturate at $\sim 70\%$--75\% above 1 keV and show a hint of a decrease towards 5 keV. Similarly, ${p}_{\mathrm{rms}}$ increases with energy in the 0.15 to $\sim 1.2$  keV range, and then decreases at higher energies.

While the pulse profiles presented in Figure \ref{fig:PR_5bin} do not explicitly show all the energy ranges discussed below, it is important to note that the analysis includes a broader range of energy bands to explore the pulsar's emission behavior comprehensively.

\begin{figure*}
	\includegraphics[width=1.0\columnwidth]{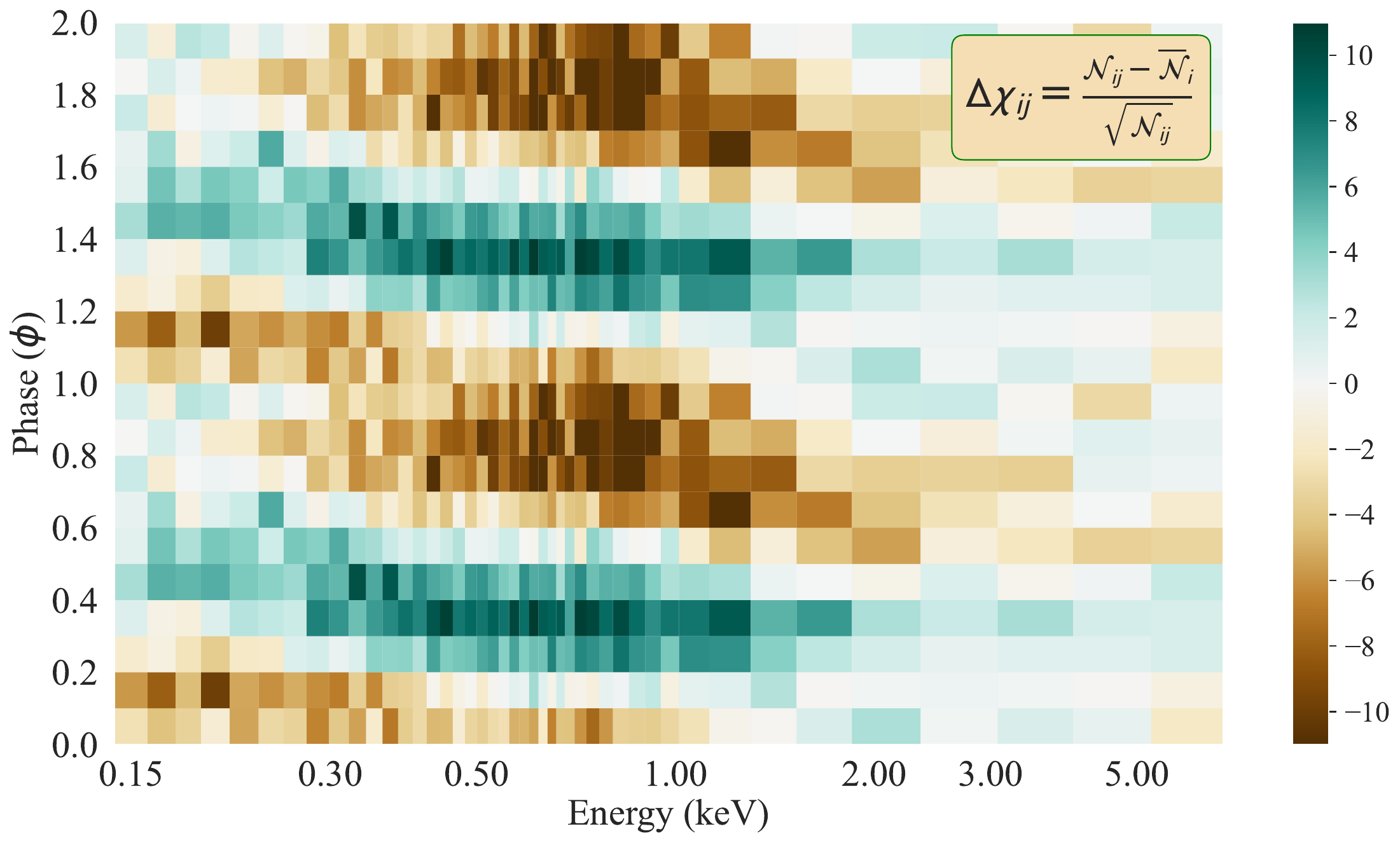}
    \includegraphics[width=1.0\columnwidth]{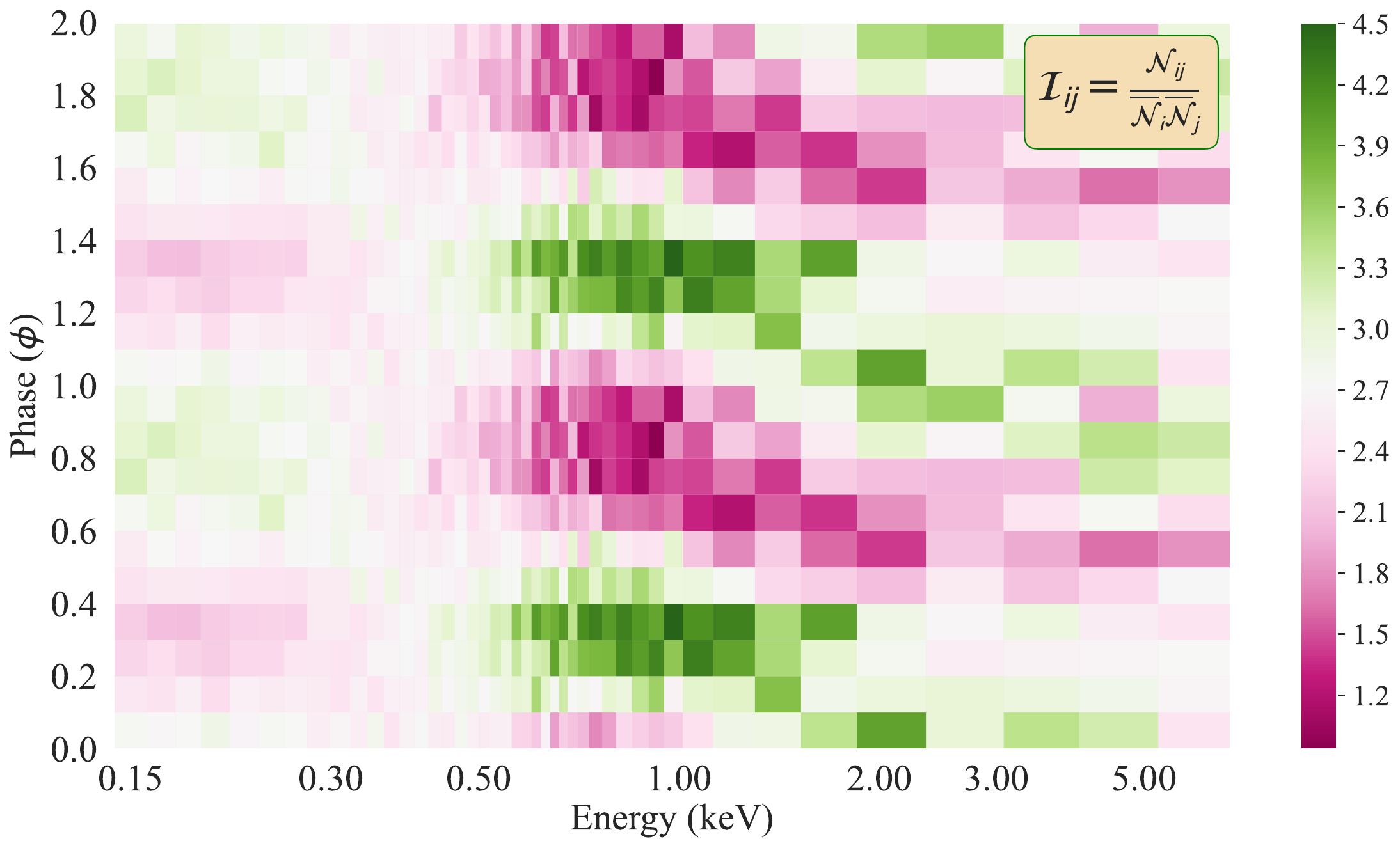}
    \caption{
    Different representations of the EPIC-pn X-ray events of \psr in phase-energy space. The right panel shows a phase-energy map where only the normalization is applied, while the left panel shows deviations from the phase-averaged values in each phase-energy bin -- see Equations (\ref{eq:map1}) and (\ref{eq:map2}). }
    \label{fig:pemap}
\end{figure*}

In accordance with Figure \ref{fig:PR_5bin}, the pulse profile at the lower energies, $E \lesssim 0.4$ keV, displays the lowest maxima and minima.

The energy range of $0.6-0.8$\,keV, where the transition between the cold and hot BB components takes place (see Figure \ref{fig:PIS_12}), exhibits the highest maxima and deepest minima. Similarly, the 1.2--2 keV corresponds to the transition region between the hot BB and PL components, with the maxima aligned in phase, but the minimum at $1.2-2$ keV occurring approximately 0.2 phases earlier.

A significant phase difference of about 0.5 is observed between the maxima and minima in the $0.45-0.6$\,keV (cold BB), $0.6-0.8$\,keV (transition from cold to hot BB), and $0.8-1.2$\,keV (hot BB) ranges. This suggests the presence of a hotter region on the NS's surface once per period, with the closest approach to the line of sight occurring at phases of $0.15-0.2$.

Lastly, in the $2-5$\,keV range (PL), the light curve seemingly exhibits two maxima and two minima per period. These maxima occur at phases around 0.2 and 0.7.

There is a noticeable difference in the shape and amplitude of the light curve between $0.15-0.3$\,keV (red color with a maximum amplitude at $\sim$ ${\cal F}(\phi) \sim$ 1.15)  and other bands.
It is noteworthy that about half ($N = 81103$) of the total events ($N = 167600$) are confined in $0.15-0.3$\,keV.\\

\begin{figure*}
    \centering
	\includegraphics[width=1.0\columnwidth]{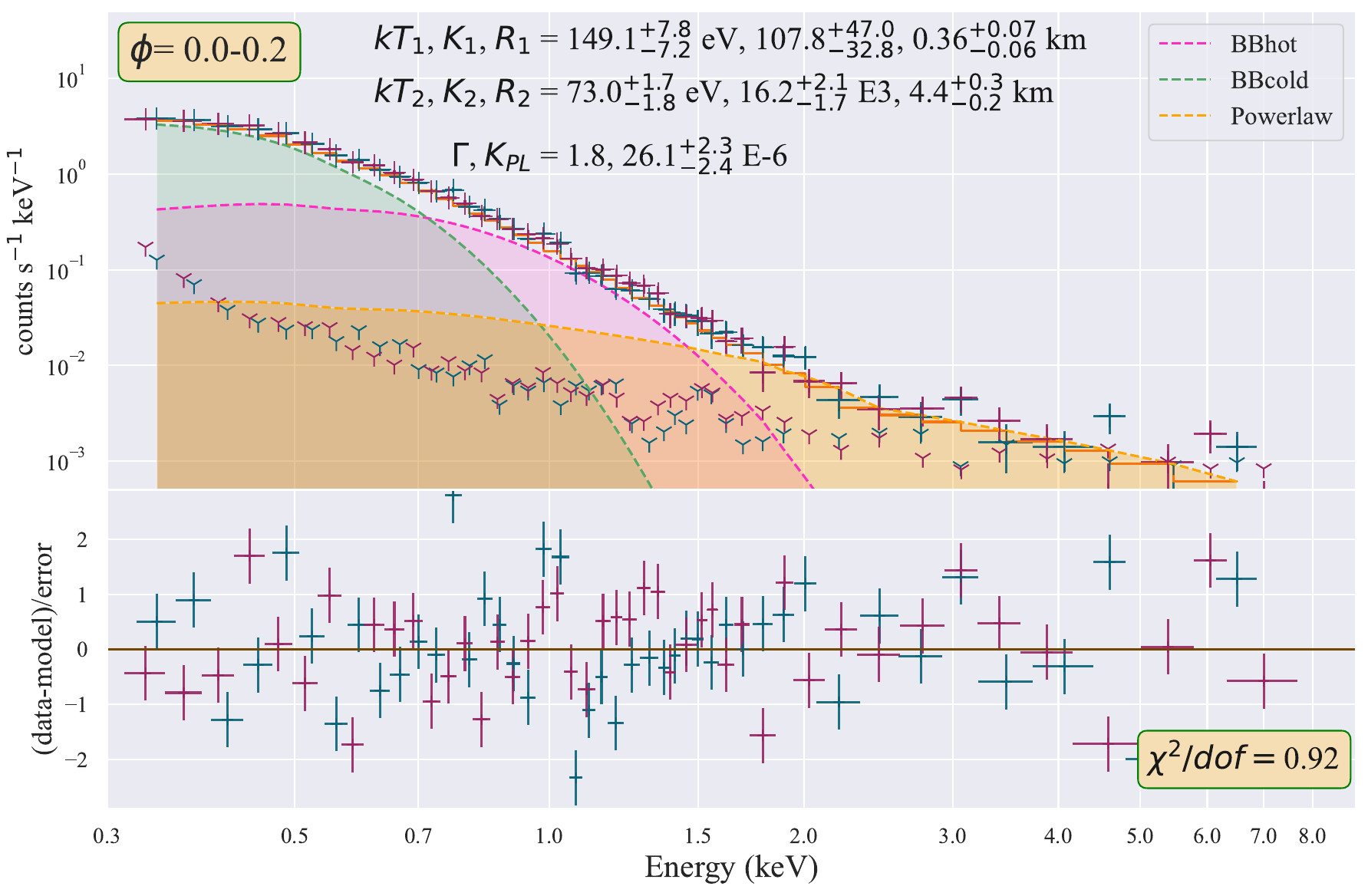}
    \includegraphics[width=1.0\columnwidth]{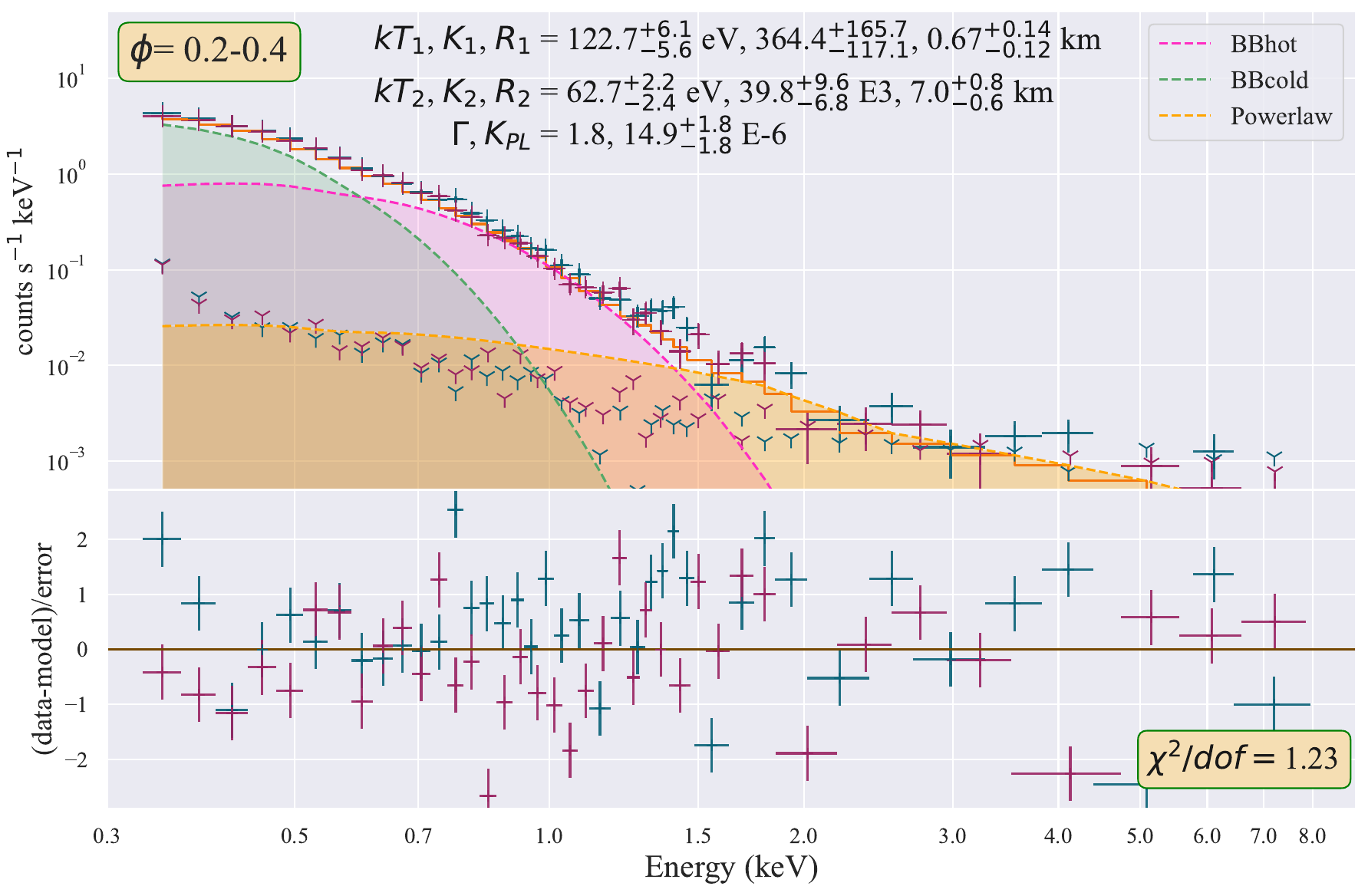}
    \includegraphics[width=1.0\columnwidth]{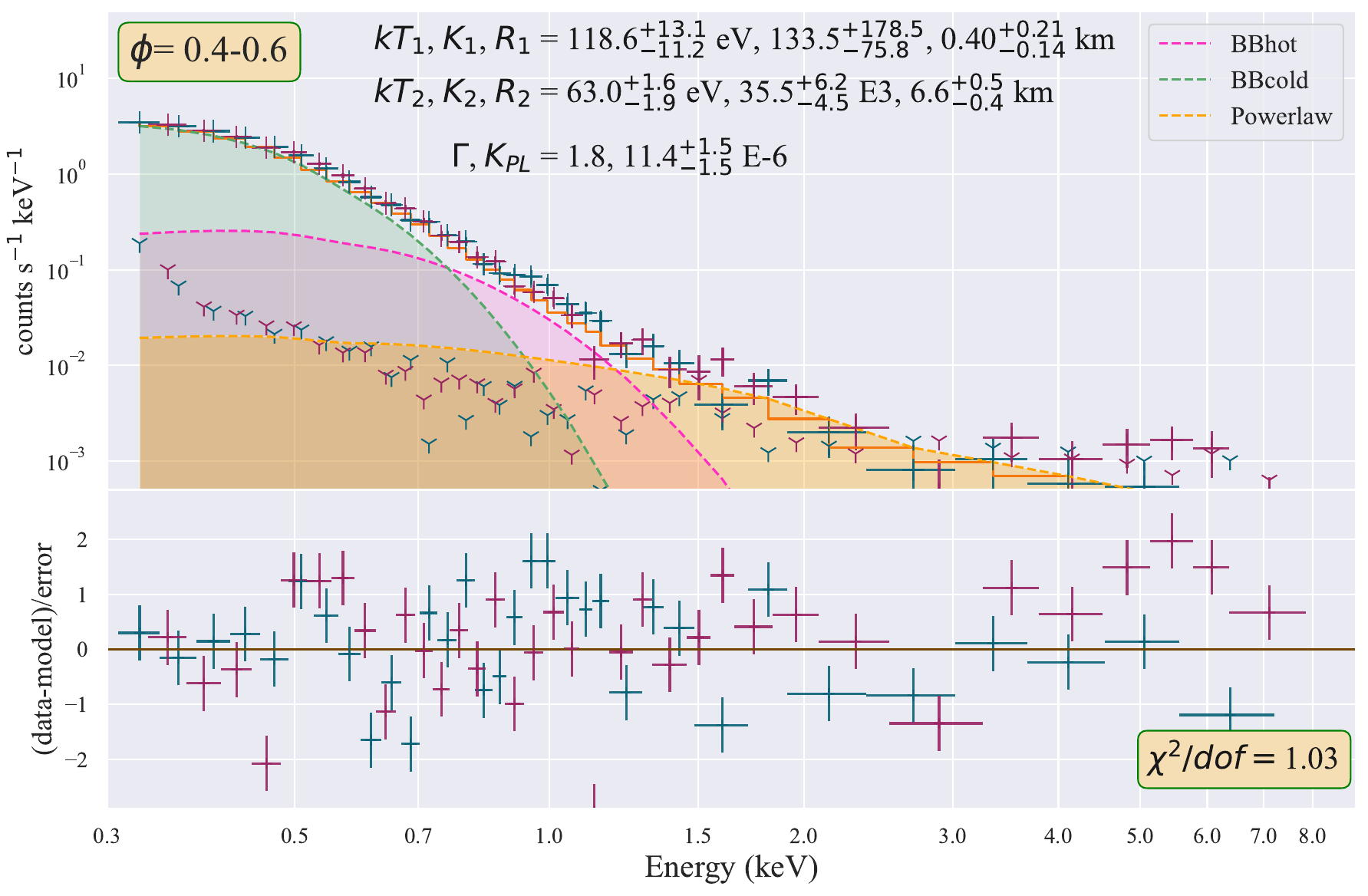}
    \includegraphics[width=1.0\columnwidth]{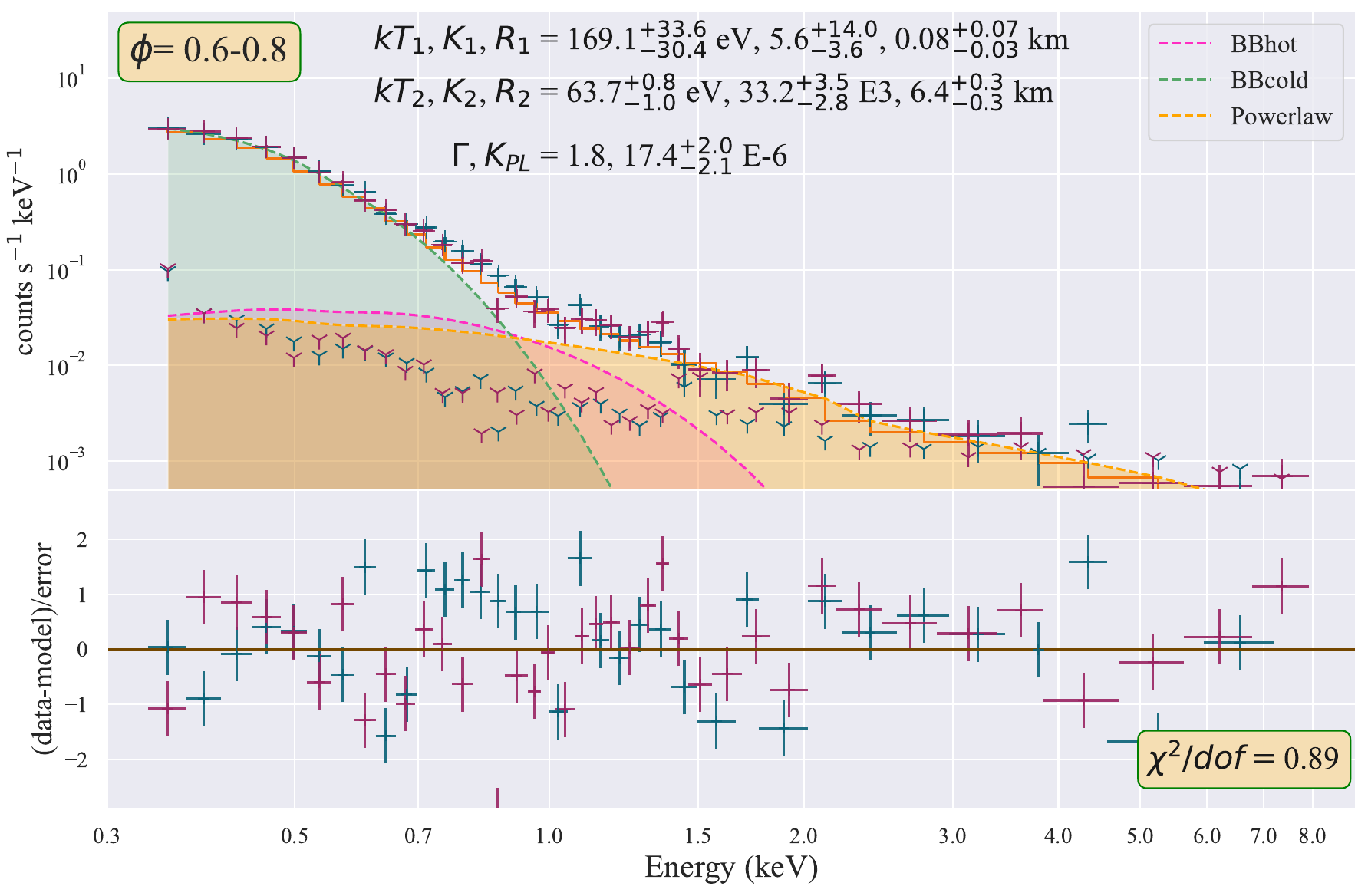}
    \includegraphics[width=1.0\columnwidth]{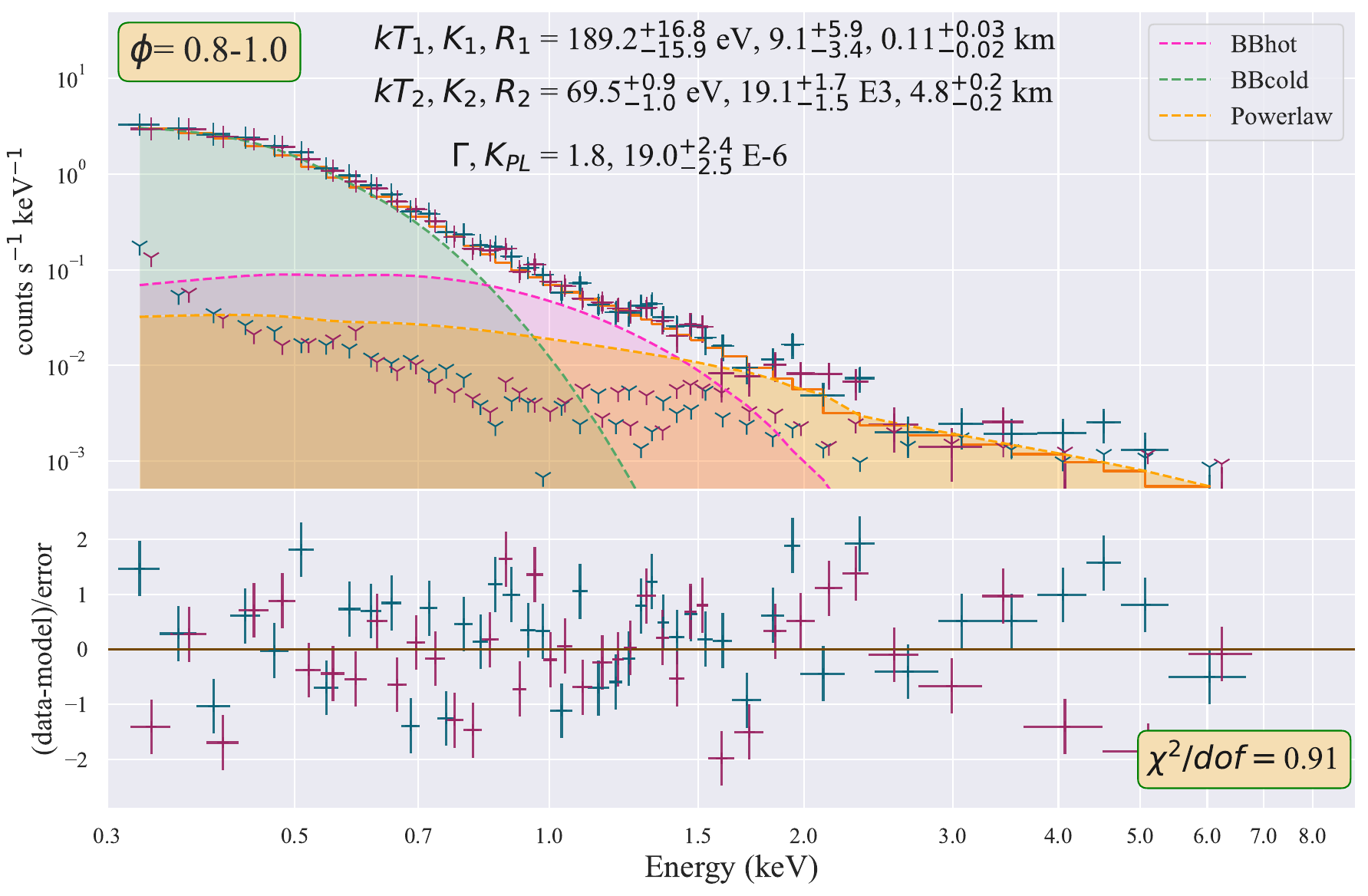}
    \caption{Fits to phase-resolved spectra for 5 equal-sized phase bins of the 2019 EPIC-pn data. Hot BB, cold BB, and PL components are displayed with red, green, and yellow colors. The fit parameters with a margin of 1\sig error are shown at the top of each panel.}
    \label{fig:phrs}
\end{figure*}

We also investigated the pulsation behavior through two different representations of the 2D phase-energy space of the X-ray events.
Following the work of \cite{Arumugasamy2018ApJ}, we calculated the deviations of the counts, $N_{ij}$ in each phase-energy bin from the phase-averaged value. These deviations from the phase-averaged values are then represented by the significance map with pixel values:
\begin{eqnarray}\label{eq:map1}
\Delta\chi_{i j} = \frac{N_{i j} -  \overline{N}_{i}}{\sqrt{N_{i j}}},
\end{eqnarray}
where $i$ and $j$ enumerate the energy and phase intervals, respectively,  and $\overline{N}_i=J^{-1}\sum_{j=1}^J N_{i,j}$ is the phase-averaged counts in the $i^{\rm th}$ energy bin, and $J$ is the number of phase bins. Here we have binned the events in the phase-energy space, using $J=10$ equal-sized phase bins and choosing variable-size energy bins to maintain $\gtrsim 40$ counts per phase-energy bin. Note that this phase-energy deviation map is restricted to the $0.15-7.0$\,keV energy range, where we have enough source counts (Figure \ref{fig:pemap} left).

Using alternatively a normalization method by \citet{Tiengo2013Natur}, the numbers of counts in phase-energy bins, $N_{i, j}$,  are divided first by the phase-averaged counts in the $i^{\rm th}$ energy bin, $\overline{N}_{i}$, and then by the energy-averaged (for the $0.15-7.0$\,keV energy band) counts in the $j^{\rm th}$ phase bin, $\overline{N}_j = I^{-1}\sum_{i=1}^I N_{ij}$, where $I$ is the number of energy channels. The resulting normalized phase-energy map has the following pixel values:

\begin{eqnarray}\label{eq:map2}
I_{ij} = \frac{N_{i j}}{\overline{N}_{i}  \overline{N}_j}
\end{eqnarray}
These results are displayed in Figure \ref{fig:pemap} (right).\\

\begin{figure*}
	\includegraphics[width=1.0\columnwidth]{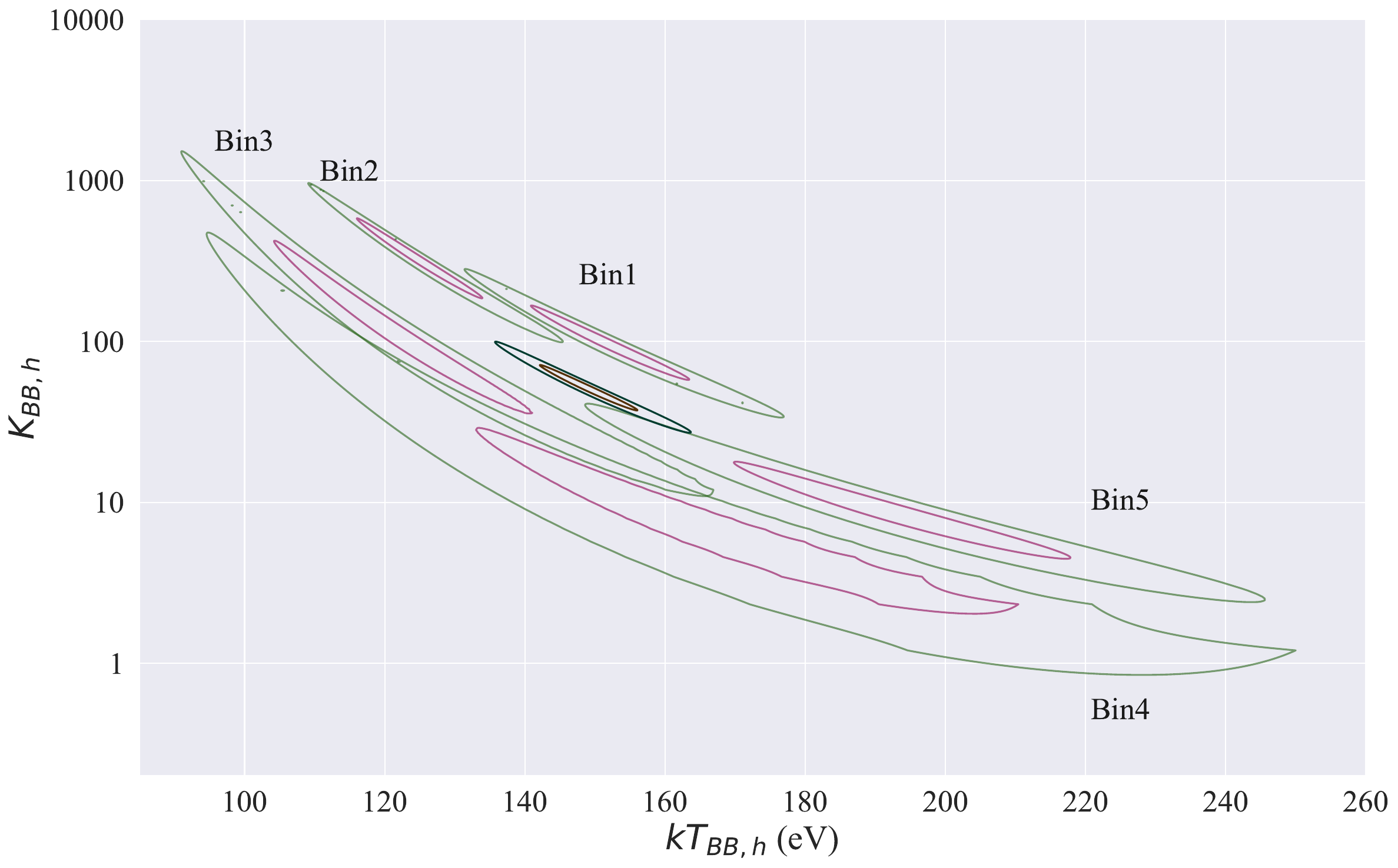}
    \includegraphics[width=1.0\columnwidth]{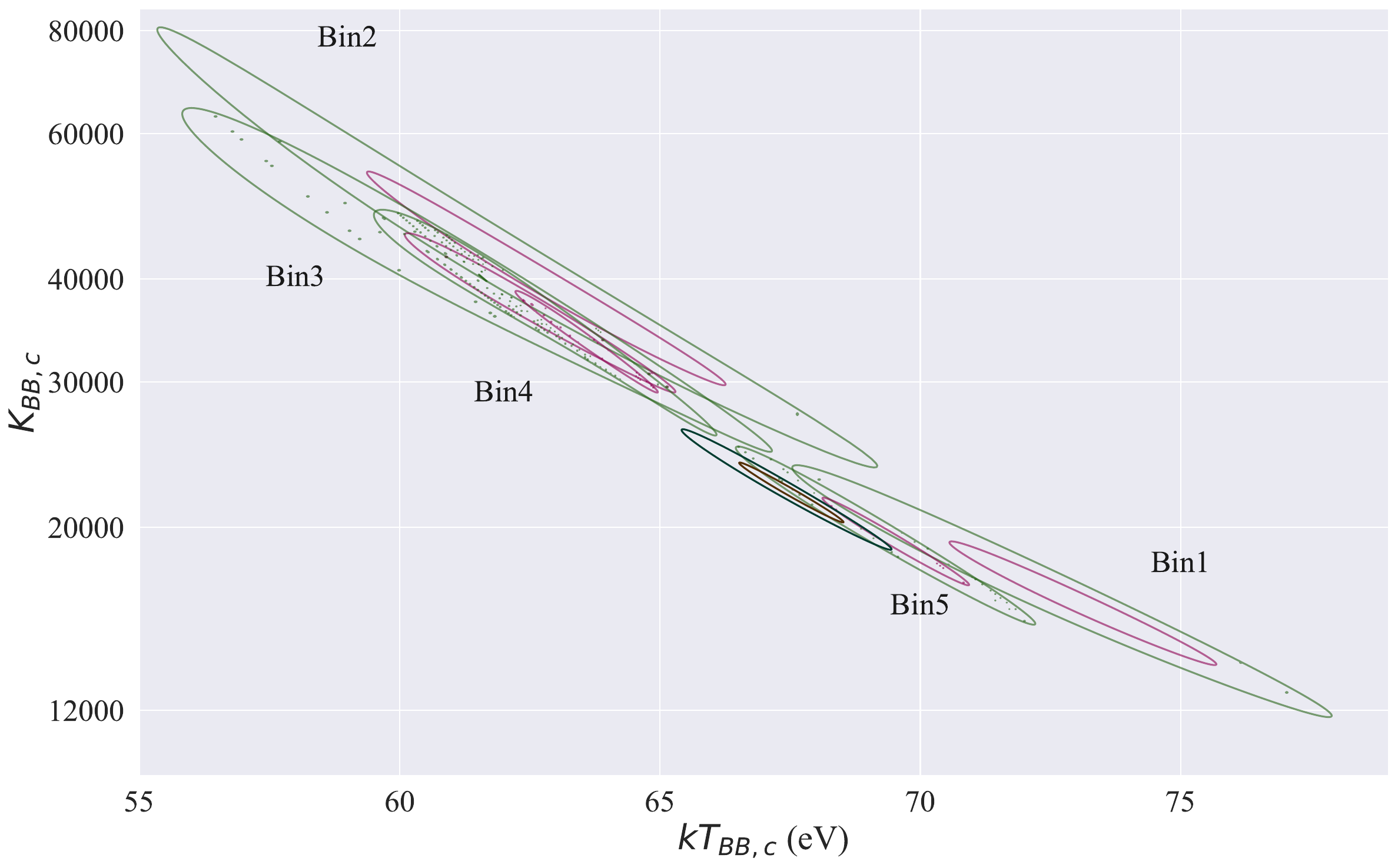}
    \caption{Variations of hot BB (left) and cold BB (right) with  phase in the temperature-normalization plane. Red and green contours correspond to confidences levels of 68.3\% and and 99\% respectively. Additionally, we depict the confidence contours of phase-integrated spectra using brown (68.3\%) and black (99\%) colors. Errors are estimated  for two parameters of interest. Small dots in some contour plots are small contours themselves representing local minima, indicating a complicated shape of the $\chi^2$ surface.}
    \label{fig:contevol}
\end{figure*}

\begin{figure*}
	\includegraphics[width=1.0\columnwidth]{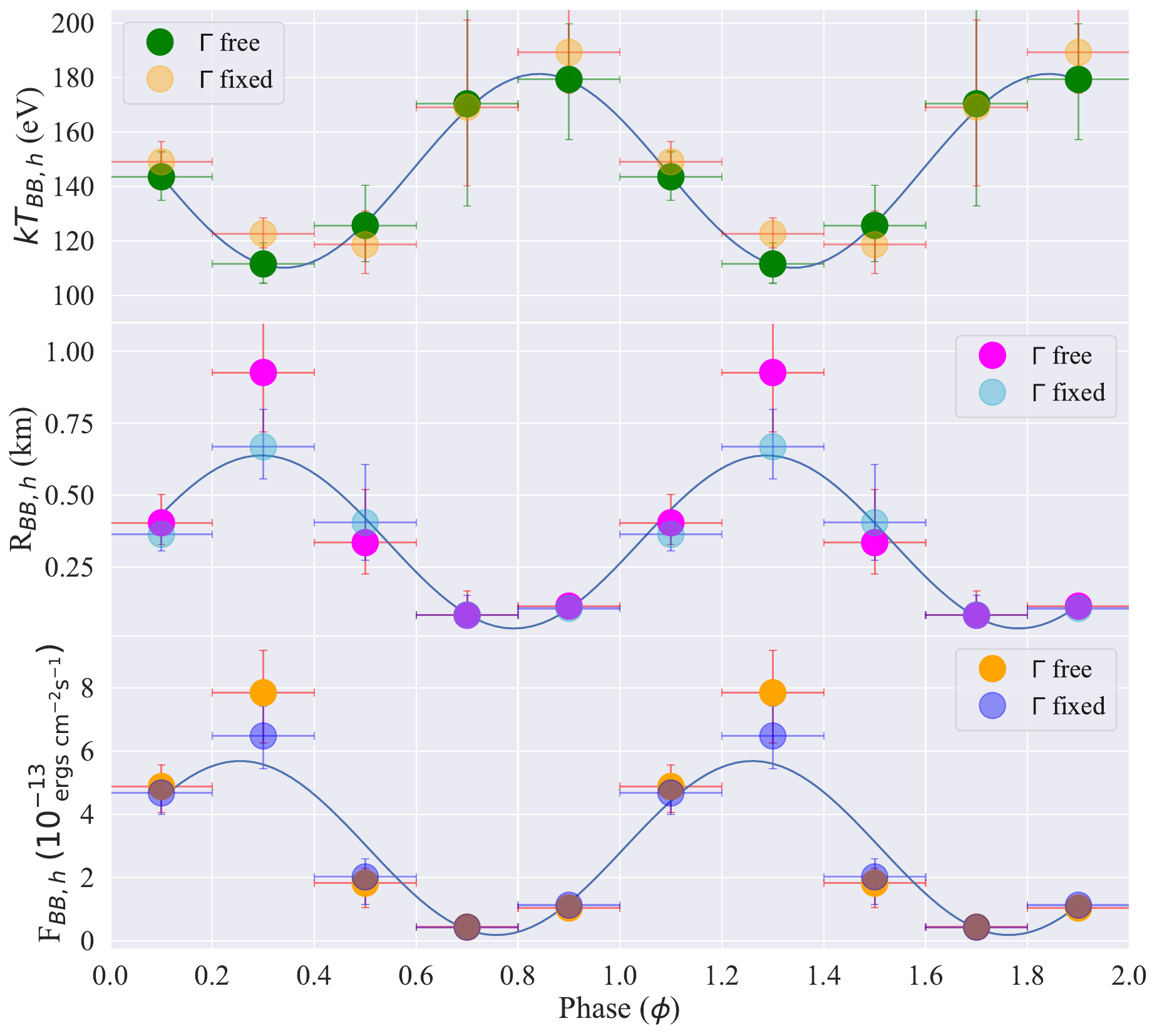}
    \includegraphics[width=1.0\columnwidth]{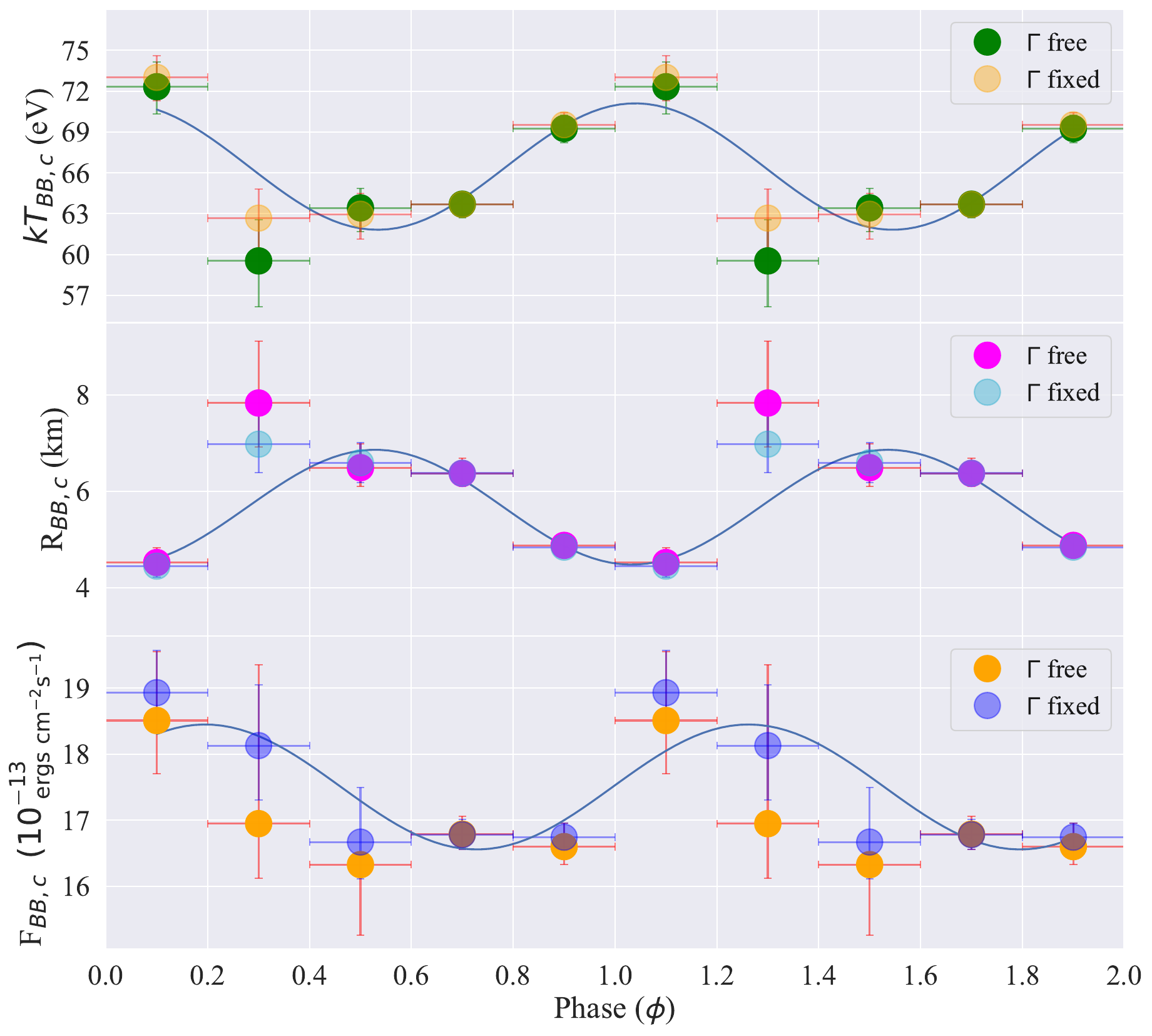}
    \caption{Phase dependencies of the temperatures (top), radii of equivalent emitting sphere  (middle), and unabsorbed fluxes in $0.3-8.0$\,keV band (bottom) for hot BB (left) and cold BB (right) components. The vertical error bars show 1\sig uncertainties of the fitting parameters. Blue lines represent sine function fits to the data for visualization purposes.}
    \label{fig:parevol}
\end{figure*}

\begin{figure}
	\includegraphics[width=1.0\columnwidth]{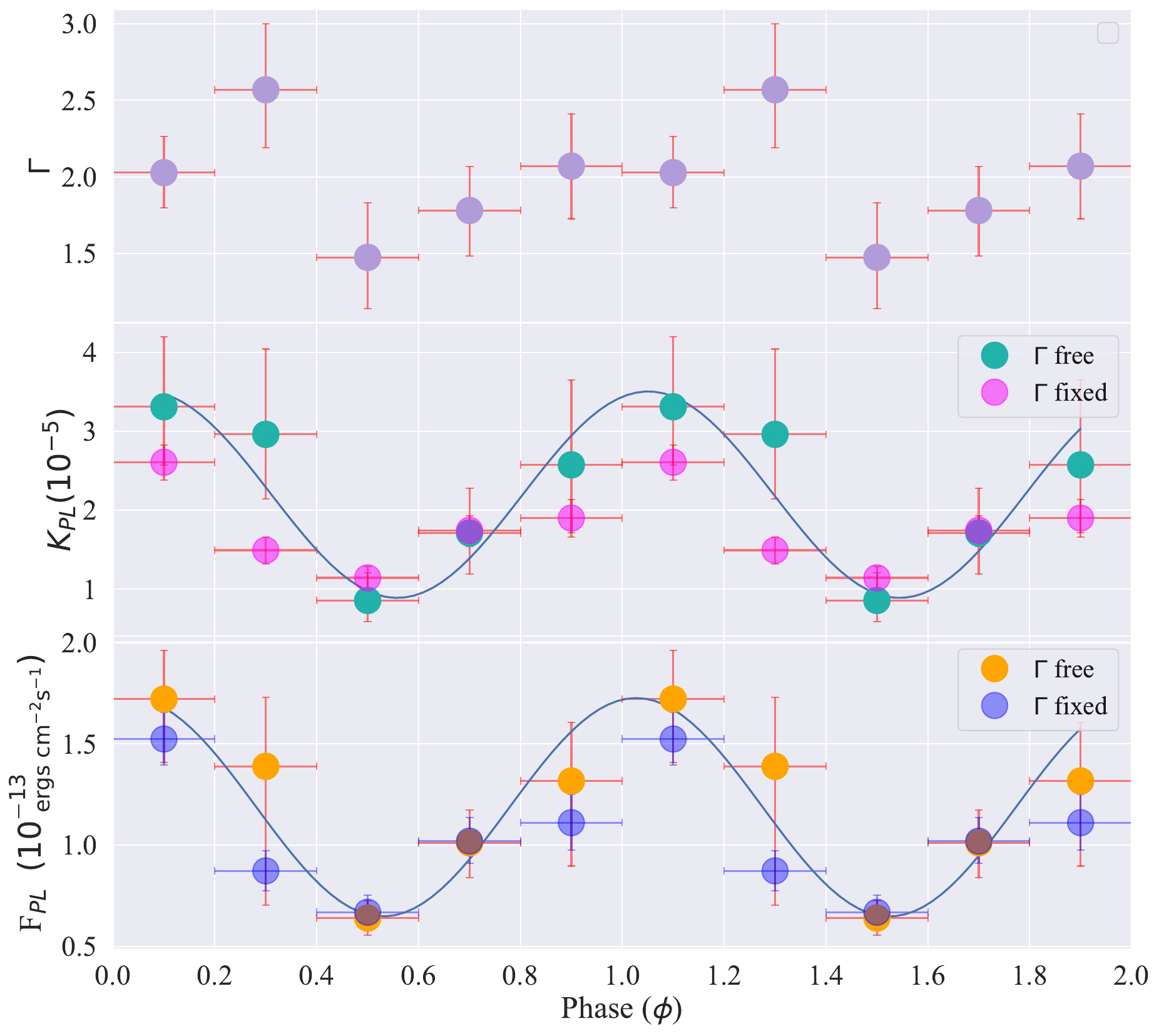}
    \caption{Phase dependence of the photon index $\Gamma$ (top), normalization $K_{\text {PL}}$ (middle), and unabsorbed fluxes in $0.3-8.0$\,keV band (bottom) for the PL component. The vertical error bars show 1\sig uncertainties of the fitting parameters. Blue lines represent sine function fits to the data for visualization purposes.}
    \label{fig:plnormevol}
\end{figure}

Upon examining the right panel (the normalized phase-energy map according to \citealt{Tiengo2013Natur}), we can classify the counts in the map into four distinct regions based on energy. In the first region ($0.15-0.3$\,keV), a minimum pulsation between phases $0.1-0.4$  is observed, followed by a faint peak between phases $0.7-0.9$.

The second region ($0.3-0.5$\,keV) shows a minimal to negligible pulsation across all phases. Interestingly, this contrasts with both the left panel (the phase-energy deviation map) and the pulse profile in Figure \ref{fig:PR_5bin}, where continuous pulsation is more evident throughout the soft to mid X-ray range.

In the fourth region ($0.5-1.5$\,keV), a pronounced pulsation maximum is present between phases $\phi$ = $0.1-0.4$, and a pulse minimum between phases $\phi$ = $0.6-1$, both of which are corroborated by Figure \ref{fig:PR_5bin}.

The count distribution in the last region ($2-5$\,keV) appears dispersed in both the right and left panels, posing a challenge in determining the nature of the observed pulsation. However, it is worth noting that a pulsation minimum and maximum are observed between $2-3$\,keV.

\section{Phase-resolved Spectral Analysis}
\label{sec:phres}

We defined 5 equal-sized phase bins and extracted spectra in each bin using the same extraction radius, filtering, and binning criteria as those in the phase-integrated analysis (Section \ref{sec:phint}). We verified that these bins contained enough (at least 15,000) counts to sufficiently constrain the spectrum. Building on the phase-integrated best 2BB+PL fit, we carry out spectral fits, freezing the absorbing hydrogen column and photon index to their best-fit values,  $N_{\text H}=2\times 10^{20}$ cm$^{-2}$ and $\Gamma = 1.8$. Each spectrum, its best-fit model parameters, and $\chi^2_{\upsilon}$ values are shown in Figure \ref{fig:phrs}. In contrast to the approach by \cite{DeLuca2005ApJ}, where the temperatures were fixed due to a low S/N ratio, our analysis benefits from a higher S/N ratio, which allowed us to vary both the temperatures and their corresponding normalizations in our spectral fit. For each bin, a good fit was obtained, constraining  the parameters of the three spectral components throughout all phases. However, in the fourth bin ($\phi$ = 0.6--0.8), the \texttt{ftest} indicated that the inclusion of the Hot BB component is statistically not required.\\
Variations of hot and cold thermal components with rotational phase are demonstrated by the temperature-normalization confidence contours for 5 phase bins (Figure~\ref{fig:contevol}), as well as the phase dependencies of  temperature, radius and unabsorbed flux (Figure~\ref{fig:parevol}).  

For the Cold BB component,  the phase-integrated spectrum yields the best-fit temperature $kT_{\rm BB,c}=69$ eV and an equivalent emitting sphere radius of $R_{\rm BB,c} =0.035 K_{\rm BB,c}^{1/2} d_{350} \sim 5$  km. We see that the best-fit temperature varies from 75 eV in Bin 1 to approximately 65 eV in Bins 2 and 3, subsequently returning to a value of 70 eV in Bin 5. The apparent radius exhibits variations anti-correlated with temperature, increasing from 4.4 km in Bin 1 to around 7 km in Bin 2 and then decreasing to 4.8 km in Bin 5. Although the  temperature-normalization contours  move almost along the direction of their maximal stretch, caused by the anticorrelation between the temperature and radius, the variation is likely real as demonstrated by the parameter uncertainties in Figure~\ref{fig:parevol}. In addition, the 99\% confidence contours of Bin 1 and Bin 5 are clearly separated from the respective contours of the other bins in Figure~\ref{fig:contevol}. 

The variability in the variations of unabsorbed $0.3-8$\,keV flux
$F_{\rm BB,c}^{\rm unab}$ exhibits a pattern akin to that of varies approximately in phase with $kT_{\rm BB,c}$ (perhaps with a slight phase shift) because the flux in the Wien tail of a thermal spectrum is particularly sensitive to temperature variations. However, the relative variations of the flux are small, within $\sim 10\%$  of its average value, because the effect of temperature variations is partly compensated by antiphase variations of emitting area, $\propto R^2$.
 
 The hot BB component exhibits more pronounced phase variations. 
 The hot temperature oscillates with phase, with minimum in Bin 2 and maximum in Bin 5, in concert with the cold BB temperature. The variations of radius (and emitting area) are anti-correlated with the temperature variations, similar to the cold BB. In contrast to the cold BB component, the unabsorbed flux associated with the hot BB component ($F_{\rm BB,h}^{\rm unab}$) varies in phase with $R_{\rm BB,h}$ (hence with the projected area of the hot region), and in antiphase with hot and cold temperatures and the cold BB flux. Such behavior is consistent with the phase shift between the low-energy and mid-energy light curves (Figure \ref{fig:PR_5bin}). The hot BB flux is substantially lower (by a factor of about 6, on average) than the cold BB flux, and it shows much stronger variations, with a minimum value close to zero, and relative amplitude of about 50\%.\\

We also checked whether allowing $\Gamma$ to vary would change the variations  thermal parameters. The results are also plotted in Figure~\ref{fig:parevol}, showing no significant differences between the cases of free and fixed $\Gamma$. For both fixed and free $\Gamma$, the PL normalization varies in antiphase with the emitting area of the hot BB component (Figure \ref{fig:plnormevol}).

\section{Discussion}
\label{sec:disc}

\subsection{Discrepancy between X-ray and Radio Timing Solutions of B1055-52}
\label{subsec:radx}

As described in Section \ref{subsec:timingsol}, we find a relative difference of the measured X-ray to the radio period of $\triangle P /{P_{\rm radio}} = 1.4 \times 10^{-8}$. Although this is in agreement with the  relative timing accuracy in different \xmm instrument modes \citep{Martin2012}, the high statistical significance of this difference prompted us to also consider a pulsar glitch as a potential non-technical explanation.

However, the simultaneous radio observations from Parkes, which are sensitive to frequency variations, show no indication of a glitch during the observed period. There is also no indication of a glitch from the three considered epochs of MeerKAT data. Moreover \psr is not known to exhibit glitching behavior. From a monthly monitoring program with Parkes since the beginning of 2007, \citet{Lower2021} reported 
only a 90\% upper limit on the size of undetected glitches of $\triangle \nu_g^{\rm 90\%} / \nu  < 3.4 \times 10^{-9}$ for B1055. This glitch limit is below our X-ray/radio discrepancy. For these reasons, we regard it as highly unlikely that B1055 experienced an undetected glitch between the X-ray observations or after the simultaneous Parkes coverage ended in the second \xmm observation.

\subsection{The phase-resolved thermal X-ray spectra}

One or two BB components are commonly used to describe the thermal spectra of  middle-aged pulsars (e.g., \citealt{caraveo2004phase, DeLuca2005ApJ} for the Three Musketeers, \citealt{Schwope2022} for PSR B0656+14; \citealt{Rigoselli2022MNRAS} for PSR 1740+1000). However, our phase-resolved spectral analysis shows that such a two-temperature model is not a fully consistent description of the data because both the hot and cold BB component show significant variations with phase. Hence, the temperature is not uniformly distributed, neither in the hot spot nor in the bulk NS surface. \cite{Arumugasamy2018ApJ} showed similar phase variations of the cold and possibly the hot BB temperatures for PSR B0656+14 (their Figure 13), although without providing 2D confidence contours for normalization and temperature.\\ 

\label{subsec:tempdist}

Based on the phase-resolved spectra of B1055 (Figure \ref{fig:phrs}) and the respective parameter evolution over phase (left panels of Figures \ref{fig:contevol} and \ref{fig:parevol}), the behavior of the hot spot BB (``$S1$'') can be summarised as follows:

Starting from Bin 1, which includes the maximum of the 0.5--1.5 keV  light curve (where the hot BB component dominates; see Figure \ref{fig:PR_5bin}), the temperature of the hot BB component decreases with increasing phase and reaches a minimum near the boundary between Bin 2 and Bin 3, i.e., on the descending part of the light curve.
While the light curve reaches its minimum near the boundary between Bin 3 and Bin 4, the temperature grows  up to its maximum at Bin 5, at the ascending part of the light curve. 
The phase shift between the temperature and the $0.5-1.5$\,keV flux oscillations is caused by the normalization (projected area) varying in opposite direction  with respect to the temperature variation. In particular, the temperature increase between Bin 3 and Bin 5 is partly compensated by the normalization decrease between Bin 2 and Bin 5, so that the increase of the flux is substantially smaller than it would be at a constant normalization.\\

The cold BB component exhibits  similar behavior to the hot BB component between Bin 1 and Bin 2, where the temperature starts to decrease and the normalization (radius) begins to increase (right panels of Figures \ref{fig:contevol} and \ref{fig:parevol}). However, during the transition from Bin 2 to Bin 4, both the $kT$ and the normalization remain nearly constant, within a 1$\sigma$ range. The cold BB component shows antiphase variations of the temperature and radius, similar to the hot BB component. \\

The observed behavior of the temperature and visible area is incompatible with the assumption of a uniform temperature in the hot and cold BB component. At this assumption the $kT$ - normalization confidence contours would be shifting along the normalization axis, in contrast to the actual observations (Figure \ref{fig:contevol}, left). 

Evidence for at least one other thermal component is also provided by the phase-energy maps. They indicate the presence of another, secondary spot (``$S2$''), best visible in the right panel of Figure \ref{fig:pemap}, at phases $\phi\sim 0.2$--0.5 and energies $E\lesssim 0.3$ keV. For a centered dipole, one would expect two similar hot spots. For a nearly-orthogonal rotator, these two hot spots should correspond to two maxima in the light curves. In principle, the location of the second hot spot could be such that only a portion of it is visible during a rotational cycle.  B1055 is likely a nearly-orthogonal rotator, as indicated by the radio pulse profile with two pulses per period (Figure \ref{fig:PR_5bin}). Therefore, second hot spots would be the natural explanation for the pulsar's $S2$ emissions. However, the $S2$'s apparent colder temperature (lower energy in the phase-energy map) is puzzling. The $S2$ is also noticeable in the light curves. In the softer energy band, $\sim 0.3-0.5$\,keV, a second smaller hump, which might be associated with $S2$, appears. It is aligned (perhaps coincidentally) with the radio inter-pulse. In addition, the phase-dependent cold BB temperature (Figure \ref{fig:parevol}) shows a flattening around the minimum ($\phi = 0.2-0.8$), possibly due to the contribution from the warm second spot $S2$.\\

Taking into account bending of photon trajectories by the  NS gravitational field and assuming locally isotropic BB emission emerging from the surface, \citet{Turolla2013} calculated light curves and the pulsed fraction, PF$_{\rm amp}$, as a function of the viewing angle and hot spot position for one and two spots with various angular radii. In particular, they found that for two antipodal spots of equal temperatures and sizes on the surface of a NS with mass $M=1.4 M_\odot$ and radius $R=15$ km, the maximum pulsed fraction, PF$_{\rm amp}\approx 29\%$ for energy-integrated flux, is reached for an orthogonal rotator. The measured PF in the $0.8-1.2$\,keV range for B1055, corresponding to the hot spot emission, is significantly higher, reaching approximately 65\%--70\%. But \citet{Turolla2013} also showed that much higher PFs (up to 100\%) can be obtained if there is only one spot, or two rather different non-antipodal spots. 
However, this finding alone does not allow us to conclude that B1055 has very different or non-antipodal hotspots because another reason for high PFs can be the presence of a light-element (H or He) NS atmosphere. Emission from such an atmosphere is beamed along the magnetic field \citep{pavlov1994}, and we can expect a higher PF if the hot spot is associated with the magnetic pole, where the magnetic field is normal to the emitting surface. We note that the (badly fitting) atmosphere models mentioned in Section~\ref{subsec:specfit} are not applicable to phase-resolved  spectra.\\

\begin{figure}
    \includegraphics[width=1.0\columnwidth]{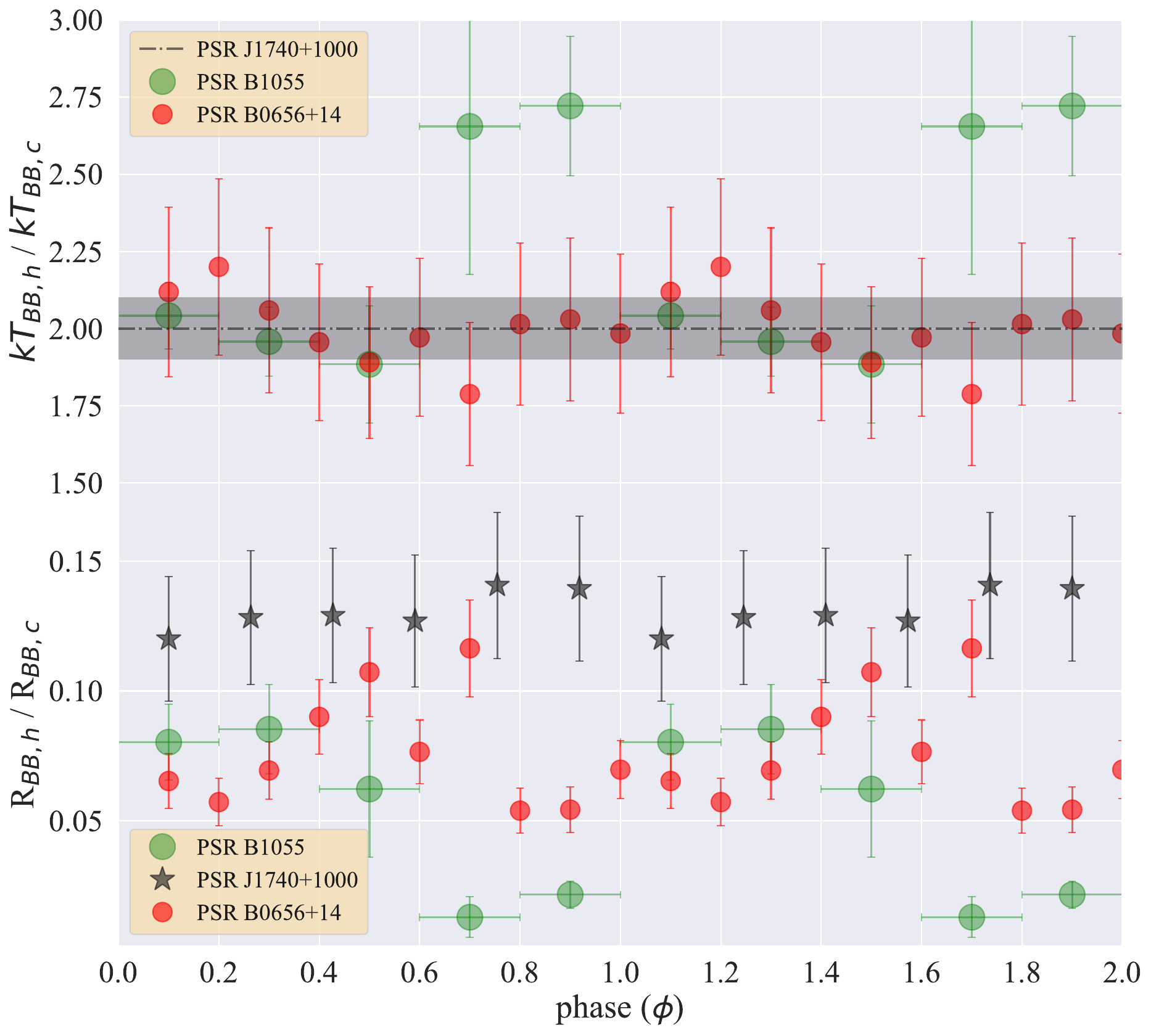}
    \caption{Phase dependencies of the temperature and radius ratios derived from 2BB fits for B1055 (this work),  PSR J1740+1000 \citep{Rigoselli2022MNRAS} and PSR B0656+14 \citep{Arumugasamy2018ApJ}. The black horizontal line in the top panel represents the phase-integrated value for PSR J1740+1000.}
    \label{fig:TRration}
\end{figure}

Hot spots could be formed by two different mechanisms. Firstly, 
they could be due to {\em anisotropic internal heating}, i.e., anisotropic  heat transfer from the very hot NS interiors to colder surface layers in the presence of a nonuniform magnetic field. 
Even a moderately strong dipole magnetic field, such as B1055's $B\sim 10^{12}$ G, can lead to noticeable temperature nonuniformity, with a colder equatorial belt and a hotter area of the rest of the surface, slowly increasing towards the magnetic poles of a centered dipole (e.g., \citealt{yakovlev2021}). However, although the spectrum of such a NS can be fitted with a BB+BB model, the ratios of hot and cold temperatures and radii are very different from the observed ones, for B1055 and other middle-aged pulsars (Figure \ref{fig:TRration}).

If, in addition to the dipolar magnetic field, there is a toroidal field component in the NS crust, then the two hot spots around the poles of the dipolar component, can have different temperature and sizes \citep{Geppert2006}. The toroidal component leads to the blanketing effect, channeling heat flow along the polar axis and creating an extended cold equatorial belt with an asymmetric temperature distribution that covers most of the stellar surface.\\ 

Any scenario that involves hot spot(s) should explain the observed high PFs. \citet{Geppert2006} estimated PFs for various surface temperature distributions involving two warm spots, assuming semi-isotropic BB emission. They obtained a maximum pulsed fraction of ${\rm PF}_{\rm amp} = 33\%$ for an orthogonal rotator, which remains much below the observed values for B1055 (65\%--70\% in the $0.8-1.2$\,keV energy range and even 38\% for energies $\lesssim 0.5$ keV).

\cite{Perna2013MNRAS} considered coupled thermal and magnetic evolution of a NS and computed surface temperature and magnetic field distributions at different ages for a variety of initial magnetic field configurations. Assuming local BB spectra, they considered not only the (semi-)isotropic angular distribution of emitted radiation but also radiation moderately beamed along the local magnetic field direction,  mimicking emission from a light-element atmosphere. Interestingly, addition of a large-scale toroidal magnetic component disrupts the north-south symmetry of the temperature distribution due to the Hall effect, resulting in single-peaked light curves and notably heightened PFs. For specific models, these PFs exceed 50\%, within the $0.5-2$\,keV energy band. Overall, the presence of a dominant large-scale toroidal magnetic component and/or atmospheric beaming effects could explain high PFs. Conversely, multipolar configurations yield intricate temperature profiles with relatively smaller PFs \citep{Perna2013MNRAS}.

For B1055, a 50\% PF is still too low. Only in combination with atmospheric effects could internal heating (in the presence of a toroidal magnetic field component) explain the large observed PFs.\\

Hot spots could be also due to {\em external heating} by relativistic particles accelerated in the pulsar's magnetosphere, which subsequently precipitate onto the NS polar caps (PCs) and heat them \citep{Ruderman1975ApJ}. This mechanism is responsible for thermal X-ray emission from hot spots in old pulsars, including millisecond pulsars, whose bulk surface is too cold to emit X-rays, but it can also contribute to the hot thermal components in middle-aged pulsars (e.g., \citealt{Pavlov2002}). The range of bolometric luminosities, $L_{\rm BB,h}$, associated with B1055's hot BB confidence contours is $\sim(3$--$5) \times 10^{30}$ erg s$^{-1}$ (Figure \ref{fig:cont}). \cite{Harding2001ApJ}, who considered the PC heating in the space charge limited flow acceleration model, predicted a polar cap luminosity $L_{\rm PC} \sim 2\times 10^{-4} \dot{E} = 6 \times 10^{30}$ erg s$^{-1}$ for a pulsar with a characteristic age $\tau \sim 500$ kyr and a spin period $P\sim 0.2$ s, consistent with our estimates of the hot BB luminosity. This suggests that the hot spots, associated with the hot thermal component in the B1055's X-ray emission, could be polar caps heated by relativistic particles precipitating from the pulsar's magnetosphere.

Furthermore, the conventional polar cap radius of B1055, calculated based on a simple ``centered'' dipole magnetic field geometry, $R_{\rm pc} = (2\pi R_{\rm NS}^3/cP)^{1/2} \approx 900$ m (assuming a standard NS radius of 13 km), is consistent with the maximum radius from the phase-resolved analysis (Figure \ref{fig:parevol}).

Yet, in this scenario, we still lack a clear explanation for the differences in heating between the two magnetic poles. One plausible explanation for the anisotropy in external heating could be associated with off-centered dipole magnetic field configurations (e.g., \citealt{harrison1975}). A displacement of the dipole center from the NS center may lead to differences for internal and external heating, thus different pole temperatures.

The effect of an off-centered dipole magnetic field in the case of internal heating was recently studied by \cite{Igoshev2023MNRAS}.
They investigated how these configurations affect the temperature patterns on the surface, the light curves, and the spectra of middle-aged pulsars.

For an off-centered dipole field with $B=10^{13}$\,G, \cite{Igoshev2023MNRAS} reported $kT_{\rm BB,h}$/$kT_{\rm BB,c} \approx 1.1$ and $R_{\rm BB,h}$/$R_{\rm BB,c} \approx 2$. These values are notably different from the typical phase-integrated values of $kT_{\rm BB,h}$/$kT_{\rm BB,c} \approx 2$ and $R_{\rm BB,h}$/$R_{\rm BB,c} \approx 0.1$ for middle aged pulsars (Figure \ref{fig:TRration}). \\

In summary, both the internal and external heating mechanisms 
can contribute to the formation of the observed hot spots on  
B1055's surface. Detailed atmosphere modeling, with account for beamed emissions, is needed for full understanding of the emission properties of middle-aged pulsars like B1055. However, such modeling is beyond the scope of this paper and is planned for future research.

\subsection{Nonthermal emission and multiwavelength pulse profile of B1055}
\label{subsec:lags}

The phase shifts between the peaks in multiwavelength pulse profiles (see Figures \ref{fig:pp_radio} and \ref{fig:PR_5bin}) can be attributed to different emission mechanisms operating at distinct emission regions that are located at different heights and azimuthal angles. For example, thermal X-ray emission comes directly from the NS surface while radio emission is thought to come from different heights above the magnetic poles. The relative shifts and shapes of the multiwavelength pulse profiles are governed by the location of the magnetic poles with respect to the rotation axis and the line of sight, in short the pulsar geometry.

\subsubsection{Nonthermal X-ray emission}

In our spectral fits with the  BB$_{\rm c}$+BB$_{\rm h}$+PL model, the nonthermal (PL) component dominates at $E\gtrsim 2$ keV (Figures \ref{fig:PIS_12} and \ref{fig:phrs}). Its phase-integrated photon index, $\Gamma_X\approx 1.8$, substantially exceeds $\Gamma_\gamma\approx 0.9$ in the GeV $\gamma$-ray range \citep{Posselt2023arXiv}. The X-ray photon index might show some variations with phase (see Figure \ref{fig:plnormevol}), but they are not statistically significant even in our relatively deep observation.

The $2-5$\,keV light curve shows one clear peak with a maximum at $\phi\approx 0.1$, i.e., within the broad $\gamma$-ray pulse (Figure \ref{fig:PR_5bin}). The maximum of the PL flux phase dependence, $F_{\rm PL}(\phi)$, is seen at about the same phase (Figure \ref{fig:plnormevol}). The X-ray maximum slightly lags the leading edge of the $\gamma$-ray pulse, but it is ahead of the mean phase of the $\gamma$-ray pulse as well as of the radio MP. This X-ray peak is also seen in the energy-phase map (Figure \ref{fig:pemap}, right panel). 

The $2-5$\,keV light curve also shows a hint of second peak at $\phi\approx0.8$, near the radio IP. This peak, however is not seen in the $F_{\rm PL}(\phi)$ curve, which suggests that it is due to an admixture of thermal hot component rather that to the nonthermal emission.

Thus, we cannot exclude the possibility that the nonthermal X-ray emission is connected with the $\gamma$-ray emission and perhaps with the radio MP. Unfortunately, the small number of detected nonthermal X-ray photons does not allow us to investigate this connection in more detail. \\

\subsubsection{Constraints on pulsar geometry from the radio and \texorpdfstring{$\gamma$}--ray light curves}

Similar to other radio pulsars with an IP, separated from the MP by about half a period, B1055 is usually interpreted as a nearly-orthogonal rotator, i.e., the magnetic inclination $\alpha$ (the angle between the rotation and magnetic axes) and the viewing angle $\zeta$ (between the rotation axis and the line of sight) are not strongly different from $90^\circ$. Although \cite{manchester1977} noted that such pulsars could also be interpreted as a nearly aligned hollow-cone rotators, 
B1055 is typically interpreted as a nearly-orthogonal rotator (see, e.g., \citealt{Weltevrede2009MNRAS}, and references therein),  
even though the lag of the IP behind the MP is $\Delta\phi=0.44$ (peak-to-peak) rather than 0.5. Based on analysis of swings of linear polarization angle in both the MP and IP and assuming a dipole magnetic field, \citet{Weltevrede2009MNRAS} estimate $\alpha \approx 75^\circ$, $\zeta\approx 111^\circ$ for the MP. They also suggest that the IP arises from emission formed on open field lines close to the magnetic axis at a height $\sim 700$ km above the magnetic pole, while the MP originates from field lines lying well outside the polar cap boundary beyond the null surface, and farther away from the magnetic axis, at about the same height. \\

Additional constraints on the origin of nonthermal emission can be obtained from the analysis of $\gamma$-ray light light curves. \cite{Pierbattista2015} computed emission patterns for Fermi-LAT pulsars, including B1055, for four $\gamma$-ray emission models: Polar Cap, Slot Gap, Outer Gap and One Pole Caustic, for the core plus cone radio emission model. For each of these models, they generated $\gamma$-ray and radio light curves across a parameter space defined by $\alpha$ and $\zeta$. However, none of these models provided a satisfactory explanation for both the $\gamma$-ray and radio light curves of B1055. Exceptionally for $\gamma$-ray pulsars, B1055's $\gamma$-ray pulse can be explained by the Polar Cap model but this  model does not work for the radio light curve. The Outer Gap model offered the most reasonable, although far from being perfect, description of the general characteristics of both the $\gamma$-ray and radio light curves, with values of $\alpha\sim 77^\circ$ and $\zeta\sim 87^\circ$. Overall, it seems that those models cannot provide an accurate description of both the $\gamma$-ray and radio pulsations.  

These models, as well as new models explaining pulsar $\gamma$-ray emission, are currently further developed and light curve fitting  is used to constrain these models, for instance for the Vela pulsar (e.g., \citealt{Barnard2022, Venter2017}, and references therein). However, we are not aware of a  recent work focusing on the unusual radio and $\gamma$-ray light curves of B1055.

\subsubsection{Pulsar geometry with account for X-ray pulsations}

Over the past two decades, X-ray observations of B1055 have consistently shown a single peak at energies corresponding to the hot BB component, i.e., one hot spot. However, if this pulsar is indeed an orthogonal rotator, as suggested by the radio data, one might expect to observe two peaks in the folded light curve, corresponding to two magnetic poles of a centered magnetic dipole.

In our XMM-Newton data we see not only the peak from the dominating hot spot $S1$, centered at $\phi\approx 0.3-0.4$, but also a hint of a secondary thermal X-ray peak, possibly associated with another hot spot ($S2$), which lags the main peak by approximately $\Delta\phi \approx 0.4$--0.5 (Figure \ref{fig:pemap}, right panel). The $S2$ hot spot also appears as a hump at $\phi \approx 0.8$ (near the phase of the radio interpulse) in the $0.3-0.5$\,keV light curve (Figure \ref{fig:PR_5bin}). 

Thus, the thermal X-ray light curves of B1055 are consistent with two observed magnetic poles, hence a orthogonal rotator geometry, but the poles show an asymmetry in their temperatures.\\

The radio IP and MP are separated by a distance of approximately $\sim$ $159^\circ$ ($\Delta\phi =0.44$, \citealt{Weltevrede2009MNRAS}). As illustrated in Figure \ref{fig:pp}, the peak of the MP exhibits a phase lag of 0.22\footnote{Our $0.22\pm0.02$ phase shift between the MP (at 0.67\,GHz) and X-ray is consistent with earlier results by \cite{DeLuca2005ApJ} ($0.2\pm0.05$).} with respect to the X-ray peak in the $0.3-7$\,keV, while the IP aligns with the smaller hump ($\phi = 0.7-0.8$) of the X-ray profile. 

Based on the the observed radio pulse profiles and phase dependence of the polarization angles, \cite{Weltevrede2009MNRAS} estimated emission heights and plotted a polar cap map showing sites of radio emission production projected onto the NS surface (see their Figure 7). In their cartographic representation, the IP emerges from directly above the polar region while the site of MP generation is offset from the central polar region. This spatial arrangement could potentially explain the time delay between the observed thermal X-ray pulse and the MP, as well as the alignment between the IP and $S2$. Qualitatively, the lags between the X-ray and radio pulses are also consistent with the two-pole interpretation and further supports the notion of B1055 being an orthogonal rotator. \\

The $2-5$\,keV light curve maximum closely aligns with that of the 0.5–1.5 keV light curve (see Figure \ref{fig:PR_5bin}), a phenomenon that may be attributed to a peculiar coincidence or an emission region of the non-thermal X-rays being close to the hot polar region. Conversely, the 2–5 keV maximum falls within the broader $\gamma$-ray pulse, which may indicate similar origin sites for the high-energy non-thermal emission.

The phase-energy map above 1.5\,keV seems to show a slight trend of the maxima/minima smoothly shifting to smaller phases with increasing energy. For the respective magnetospheric X-ray emitting particles, this trend may indicate different emission heights or a different angular separation from the magnetic pole. Figure  \ref{fig:PR_5bin} also shows
a statistically insignificant indication of a second hump in proximity to the location of $S2$ and the IP. This observation is in agreement with the interpretation of $S2$ as a second magnetic pole. Puzzlingly, the $\gamma$-ray light curve shows a minimum at the respective rotation phase.\\

\subsection{Absorption lines in middle-aged pulsars}
\label{subsec:comp}

Absorption features or hints for them were reported in the phase-resolved X-ray spectra of the two other Musketeers, B0656+14 \citep{Arumugasamy2018ApJ,Schwope2022} and Geminga \citep{Jackson2005}, as well as for PSR J1740+1000, another interesting middle-aged pulsar \citep{Kargaltsev2008ApJ}.  However, in later XMM-Newton observations, \cite{Rigoselli2022MNRAS} found no evidence of spectral lines in the phase-averaged and phase-resolved spectra of this pulsar suggesting a time variation in the spectrum.

An explanation for phase-dependent absorption features could be provided by multipolar magnetic field arcs that trap electrons in close proximity to the NS surface. Through the process of cyclotron resonance scattering, this may lead to the formation of phase-dependent absorption lines in the X-ray spectra (e.g, \citealt{Arumugasamy2018ApJ}). Such absorption lines can be alternatively explained by proton cyclotron lines if the local magnetic fields are high (e.g., \citealt{Tiengo2013Natur}). \cite{Vigan2014} showed that small-scale temperature variations can also mimic absorption lines.

Our analysis of the phase-resolved spectrum of B1055 did not show any evidence for absorption features, similar to the earlier study by \citet{DeLuca2005ApJ}. However,  B1055 is notably fainter than the two other Musketeers.  Considering this faintness and the variability seen for J1740+1000, the current non-detection for B1055 does not exclude the possibility of detecting phase-dependent spectral features, deeper X-ray observations achievable  with more sensitive telescopes such as the upcoming Athena X-ray Observatory.\\

\section{Conclusion}
\label{sec:conc}

Using the most comprehensive X-ray data set to date, we find that, in agreement with the findings by \cite{DeLuca2005ApJ}, the spectrum of B1055 is best described by a tripartite model consisting of two blackbody components, presumably emitted from the bulk of the NS surface and small hot spots, and a power-law component, emitted from the pulsar's magnetosphere.

A phase-resolved spectral analysis shows periodic variations in the temperature of both the hot and cold blackbody components. These variations, in tandem with periodic changes in the projected emission areas, suggest non-uniform temperature distributions both over the bulk NS surface and within the alleged hot spot(s). Previously published results on X-ray pulsations of the thermal components showed only one peak per period, likely associated with one visible hot spot, which did not agree with the nearly-orthogonal rotator geometry of B1055, implied by the detailed radio studies of B1055's main pulse and interpulse. In the new X-ray data, we find indications for a second hot spot. However, it appears to be cooler than the already known hot spot.\\

We explore two potential mechanisms to explain the thermal X-ray emission patterns of B1055: external heating by relativistic particles accelerated in the pulsar's magnetosphere, and internal heating resulting from  anisotropic heat transfer due to an offset of the dipole magnetic field and/or the presence of an additional toroidal magnetic field component within the NS's crust. Both mechanisms can, in principle, explain the observed high pulse fractions and phase dependence of the spectral parameters, particularly if beaming effects due to an atmosphere above the hot spots are additionally considered. The faintness and lower temperature of the putative second hot spot could perhaps be explained by an offset dipole, but more detailed modeling is needed.

The complexity of B1055's spectral and temporal characteristics underscores the need for further investigations, including detailed atmosphere modeling and considerations of beamed emissions.

\begin{acknowledgments}
We express our gratitude to Colin J. Clark for his help with the Fermi-LAT data, Michael Freyberg for his assistance with EPIC-pn calibration issues, Prakash Arumugasamy for his support in resolving software-related challenges, as well as to Andrei Igoshev for useful discussions. This work was supported by the Bundesministerium f{\"u}r Wirtschaft und Energie through Deutsches Zentrum f{\"u}r Luft-und Raumfahrt (DLR) under the grant number 50 OR 1917. Support for this work was also provided by the National Aeronautics and Space Administration through the XMM-Newton award No.\ 80NSSC20K0806. BP acknowledges funding from the STFC consolidated grant to Oxford Astrophysics, code ST/000488. AV thanks the LSSTC Data Science Fellowship Program, which is funded by LSSTC, NSF Cybertraining Grant \#1829740, the Brinson Foundation, and the Moore Foundation; his participation in the program has benefited this work.
\end{acknowledgments}

\vspace{5mm}
\facilities{\\
\xmm(EPIC), Parkes, MeerKAT, Fermi-LAT}

\vspace{3mm}

\software{\\
HEAsoft (v6.29c, \citealt{Hsoft}),\\ 
XMM-SAS (v20.0.0, \citealt{Gabriel2004})\\ 
XSPEC (v12.12.0, \citealt{Arnaud1996}),\\ 
PyXspec (v2.1.0, \citealt{Gordon2021}),\\ 
Astropy (v5.2.1, \citealt{astropy:2018}),\\  
NumPy (v1.23.5, \citealt{harris2020array}),\\ 
SciPy (v1.10.0, \citealt{2020NatMe}),\\
Matplotlib (v3.6.3, \citealt{Hunter:2007}),\\
Pandas (v1.5.3, \citealt{reback2020pandas}),\\
Seaborn (v0.12.2, \citealt{Waskom2021}),\\
Stingray (v1.1.1, \citealt{Huppenkothen2019ApJ}).
}



\clearpage

\bibliographystyle{aasjournal}



\end{document}